\documentclass{aa}
\usepackage{tabularx}
\usepackage{color}
\usepackage{multirow}
\usepackage{txfonts}
\usepackage{tabularx}
\usepackage{bm}
\usepackage{epstopdf}
\usepackage{subfigure}
\def\degree{\mbox{$^{\circ}$}}
\usepackage{amsmath}
\usepackage{ulem}

\usepackage{comment}

\begin{document}
\title{CHANG-ES XIV: Cosmic-ray propagation and magnetic field strengths in the radio halo of NGC~4631\thanks{Based on observations
with the 100-m telescope of the Max-Planck-Institut f\"ur Radioastronomie (MPIfR) at Effelsberg and the Karl G. Jansky Very Large Array (VLA) operated by the NRAO. The NRAO is a facility of the National Science Foundation operated under agreement by Associated Universities, Inc.}}
\author{Silvia Carolina Mora-Partiarroyo\inst{1} \thanks{\email{silvia.carolina.mora@gmail.com}} \and
Marita Krause\inst{1} \thanks{\email{mkrause@mpifr-bonn.mpg.de}}
\and
Aritra Basu\inst{2}
 \and
 Rainer Beck\inst{1}
  \and
 Theresa Wiegert\inst{3}
 \and
 Judith Irwin\inst{3}
 \and 
 Richard Henriksen\inst{3}
 \and 
 Yelena Stein\inst{4,5}
 \and
 Carlos J. Vargas\inst{13}
 \and
 Volker Heesen\inst{6}
 \and
 Ren{\'e} A. M. Walterbos\inst{7}
 \and
 Richard J. Rand\inst{8}
 \and
  George Heald\inst{9}
 \and
 Jiangtao Li\inst{10}
 \and
 Patrick Kamieneski\inst{11}
 \and
 Jayanne English\inst{12}
}
\institute{Max-Planck-Institut f\"ur Radioastronomie, Auf dem H\"ugel 69, 53121 Bonn, Germany
\and
Fakult{\"a}t f{\"u}r Physik, Universit{\"a}t Bielefeld, Postfach 100131, 33501 Bielefeld, Germany
\and
Department of Physics, Engineering Physics, \& Astronomy, Queen's University, Kingston, ON, K7L 3N6, Canada
\and
Observatoire Astronomique de Strasbourg, Universit{\'e} de Strasbourg, CNRS, UMR 7550, 11 rue de l’Universit{\'e},
67000 Strasbourg, France
\and
Astronomisches Institut (AIRUB), Ruhr-Universit{\"a}t Bochum, Universit{\"a}tsstrasse 150, 44801 Bochum, Germany
\and
Hamburger Sternwarte, Universit{\"a}t Hamburg, Gojenbergsweg 112, 21029 Hamburg, Germany
\and
Department of Astronomy, New Mexico State University, Las Cruces, NM 88001, U.S.A.
\and
Department of Physics and Astronomy, University of New Mexico, 1919 Lomas Blvd. NE, Albuquerque, NM 87131, U.S.A.
\and
CSIRO Astronomy and Space Science, PO Box 1130, Bentley, WA 6102, Australia
\and
Dept. of Astronomy, University of Michigan, 311 West Hall, 1085 S. University Ave., Ann Arbor, MI 48109, U.S.A.
\and
Dept. of Astronomy, University of Massachusetts, 710 North Pleasant St., Amherst, MA, 01003, U.S.A.
\and
Department of Physics and Astronomy, University of Manitoba, Winnipeg, Manitoba, R3T 2N2, Canada
\and
Department of Astronomy and Steward Observatory, University of Arizona, 933 N Cherry Ave, Tucson, AZ 85719 U.S.A.}
\date{Received 04.11.2018 / Accepted 17.08.2019}

\abstract{} 
{NGC~4631 is an interacting galaxy that exhibits one of the largest, gaseous halos observed among edge-on galaxies. We aim to examine the synchrotron and cosmic-ray propagation properties of its disk and halo emission with new radio continuum data.}
{Radio continuum observations of NGC~4631 were performed with the Karl G. Jansky Very Large Array at C-band (5.99~GHz) in the C and D array configurations, and at L-band (1.57~GHz) in the B, C, and D array configurations. Complementary observations of NGC~4631 with the Effelsberg telescope were performed at 1.42 and 4.85~GHz. The interferometric total intensity data were combined with the single-dish Effelsberg data in order to recover the missing large-scale total power emission. The thermal and nonthermal components of the total radio emission were separated by estimating the thermal contribution through the extinction-corrected H$\alpha$ emission. The H$\alpha$ radiation was corrected for extinction using a linear combination of the observed H$\alpha$ and 24~$\mu$m data.}
{NGC~4631 has a global thermal fraction at 5.99 (1.57)~GHz of 14$\pm$3\% (5.4$\pm$1.1\%).
The mean scale heights of the total emission in the radio halo (thick disk) at 5.99 (1.57)~GHz are $1.79\pm0.54$~kpc ($1.75\pm0.27$~kpc) and have about the same values for the synchrotron emission. The total magnetic field of NGC~4631 has a mean strength of $\rm{\langle B_{eq}\rangle} \simeq 9~\rm{\mu G}$ in the disk, and a mean strength of $\rm{\langle B_{eq}\rangle}~\simeq 7~\rm{\mu G}$ in the halo. We also studied a double-lobed background radio galaxy southwest of NGC~4631, which is an FR~II radio galaxy according to the distribution of spectral index across the lobes.}
{From the halo scale heights we estimated that the radio halo is escape-dominated with convective cosmic ray propagation, and conclude that there is a galactic wind in the halo of NGC~4631.}


\titlerunning{CHANG-ES XIV: Scaleheights and magnetic fields in the halo of NGC 4631}
\authorrunning{Mora-Partiarroyo et al.} 
\maketitle

\section{Introduction}
\label{intro}

The halo of NGC~4631 is one of the largest known for edge-on galaxies. X-ray emission from hot thermal gas has been observed up to 8~kpc above the plane \citep{Wang1995,Wang2001}. There is also evidence for dust in the halo of NGC~4631. \cite{Neininger1999} found cold dust ($\lambda$1.2~mm) up to distances larger than 10~kpc above the midplane of the galaxy, which correlates with H{\sc i} features. In addition, \cite{Alton1999} identified several chimneys or filaments at $\lambda$450~$\mu$m and 850~$\mu$m, which they interpreted as dust outflows connected to the star formation activity in NGC~4631. New observations by \cite{Melendez2015} with the {\it Herschel} infrared telescope also show a complex of filaments and chimney-like features at $\lambda$70~$\mu$m and 160~$\mu$m that extend up to 6~kpc above the plane of the galaxy. Two highly energetic supershells have been detected in H{\sc i} near the midplane with diameters of 3 and 1.8~kpc (\cite{Rand1993}. 

The central region of the galaxy has a radio structure consisting of three
collinear emission peaks \citep{Duric1982}, hereafter denoted as the "triple radio source" (see Fig. \ref{total_radio_over_optical}). The central peak coincides with the infrared (IR) center at $\rm \lambda~2.2~\mu$m \citep{Aaronson1978}. The western peak resolves into a complex of smaller structures, one of these is possibly a background radio galaxy \citep{Golla1999}.

NGC~4631 is highly disturbed in its eastern disk, as seen in dust \citep{Neininger1999}, H{\sc i} \citep{Rand1994}, CO \citep{Golla+Wielebinski, Rand2000}, H$\alpha$ \citep{Golla1996, Hoopes1999}, radio \citep{Hummel1990, Golla1994}, and optical observations \citep{Ann2011}. On the other hand, the western disk is consistent with a normal spiral viewed edge-on. The position angle of the western disk is 83\degree\ \citep{Golla+Wielebinski, Rand1994}, and around 90\degree\ along the eastern half \citep{Rand2000}. 

The radio continuum in spiral galaxies emerges from mainly two mechanisms: thermal free-free and nonthermal synchrotron emission. Synchrotron emission can be used to infer properties of the magnetic field and study the propagation of cosmic rays (CRs) in galaxies. To achieve this, the thermal component must be separated from the total radio emission. Due to the steeper power law behavior of the synchrotron spectrum as compared to that of the thermal spectrum, it is expected that the thermal fraction increases with increasing frequency. At high radio frequencies, it is therefore important to separate the thermal and nonthermal components, especially in the context of edge-on galaxies.

The thermal emission in face-on galaxies has been estimated in spatially resolved
studies (e.g., \citealt{Fatemeh2007} and \citealt{Basu2012, Basu2017}). In the case of NGC~4631, an additional complication originates from the fact that the galaxy is observed edge-on and the long lines of sight can introduce significantly more effects of dust attenuation when compared to face-on galaxies. \cite{Mora2016} compared the thermal emission of NGC~4631 on a pixel-by-pixel basis derived using three different approaches to account for dust extinction. This study has been extended to a larger sample of edge-on galaxies and methodologies to estimate the thermal emission by \cite{Vargas2018}. These authors conclude that the most promising approach to estimate the thermal emission in edge-on galaxies consists of using using H$\alpha$ emission, corrected for internal absorption by dust with 24~$\mu$m data.

In this paper we present observations of NGC~4631 with the Karl. G. Jansky Very Large Array (VLA) at C-band and L-band (Sect. \ref{VLA_observations}) and with the Effelsberg telescope at 1.42 and 4.85~GHz (Sect.~\ref{Effelsberg_observations}). The single-dish and interferometric data were combined in total power (Sect.~\ref{merging_section}). Results for total power (TP), thermal and nonthermal separation, vertical scale heights, and spectral indices of NGC~4631 are presented in Sect.~\ref{Results}. This is followed by a discussion in Sect.~\ref{Discussion} and conclusions in Sect.~\ref{Conclusions}. The parameters of NGC~4631 assumed in this study are presented in Table~\ref{NGC}. In all radio maps presented in this paper, the beam area is shown as a filled circle in the left-hand corner of each image.

\begin{table}[h]
\begin{minipage}{1\columnwidth}
\centering
\caption[Parameters of NGC~4631]{\small{Parameters of NGC~4631}}
\label{NGC}
\begin{tabularx}{1\columnwidth}{c*2{>{\centering\let\\=\tabularnewline}X}}
\hline \hline
Parameter & Value\\ \hline 
Morphological type & SBcd \\
\multirow{2}{*}{Dynamical center} & $\alpha_{2000}=12^\mathrm{h}42^\mathrm{m}08^\mathrm{s}$ \\
& $\delta_{2000}= 32^{\circ}32'29 \farcs4$\tablefootmark{a} \\  
Inclination & $89^{\circ} \pm 1^{\circ}$ \tablefootmark{b} \\
Position angle of major axis & $85^{\circ}$\tablefootmark{c} \\
Distance & 7.6~Mpc\tablefootmark{d} \\ 
\hline
\end{tabularx}
\tablefoot{
\tablefoottext{a}{IR center at 2.2~$\rm{\mu m}$ \citep{Irwin2011}.}
\tablefoottext{b}{From this paper, see Sect.~\ref{section_scale_heights}.}
\tablefoottext{c}{We assume an average position angle of 85\degree\ along the major axis.}
\tablefoottext{d}{From \cite{Seth2005}, determined from the tip of the red giant branch method. -- An angular size of $1'$ corresponds to 2.2~kpc.}}
\end{minipage}
\end{table}

\section{Observations and data reduction}
\label{Observations}

\subsection{VLA data at C-band (5.99~GHz) and L-band (1.57~GHz)}
\label{VLA_observations}

Radio polarimetric observations of NGC~4631 were performed with the Karl G. Jansky Very Large Array (VLA) during its commissioning phase through the project Continuum HAlos in Nearby Galaxies -- an EVLA Survey (CHANG-ES). The CHANG-ES project is a survey of 35 edge-on galaxies that exploits the new wide-band capabilities of the VLA \citep{Irwin2012, Irwin2012_I, Irwin2013, Wiegert2015, Irwin2015, Li2016, damas2016,Irwin2017,krause+2018,Vargas2018}. The data were taken at C-band (5.0 -- 7.0 GHz) in C and D array configuration, and at L-band (1.25 -- 1.50 and 1.65 -- 1.90~GHz) in the B, C, and D array configurations. Details of the observations are summarized in Table~\ref{table_EVLA_observations}.

\begin{table*}[h!]
\caption[Details of VLA observations.]{Details of VLA observations of NGC~4631.}
\label{table_EVLA_observations}
\vspace{1mm}
\begin{minipage}{1\textwidth}
\begin{tabularx}{1\textwidth}{c|>{\centering\arraybackslash}X|>{\centering\arraybackslash}X|>{\centering\arraybackslash}X|>{\centering\arraybackslash}X|>{\centering\arraybackslash}X} 
\hline \hline
Frequency band & \multicolumn{2}{c|}{C-band} & \multicolumn{3}{c}{L-band}\\ \hline
Array configuration & C & D & B & C & D\\ \hline
Center frequency & \multicolumn{2}{c|}{5.99~GHz}  & \multicolumn{3}{c}{1.57~GHz}\\
Bandwidth\tablefootmark{a} & \multicolumn{2}{c|}{2.0~GHz}  & \multicolumn{3}{c}{0.5~GHz}\\
Spectral resolution\tablefootmark{b} & \multicolumn{2}{c|}{4.0~MHz}  & \multicolumn{3}{c}{0.5~MHz}\\ 
Observing date & 06/04/2012 & 29/12/2011 & 03/06/2012 & 02/04/2012  & 30/12/2011\\
Pointings & \multicolumn{2}{c|}{2\tablefootmark{c}}  &  \multicolumn{3}{c}{1\tablefootmark{d}}\\
Observing time\tablefootmark{e} & 6~hrs  & 75~min  & 2~hrs & 77~min  & 18~min\\
Primary calibrator & \multicolumn{2}{c|}{3C286} &\multicolumn{3}{c}{3C286}\\
Secondary calibrator & \footnotesize{J1221+2813} & \footnotesize{J1310+3220} & \multicolumn{3}{c}{J1221+2813}\\ 
Polarization leakage calibrator  &\footnotesize{OQ208}& \footnotesize{J1310+3220} &  \multicolumn{3}{c}{OQ208}\\
Restoring beam\tablefootmark{f} & $2\farcs6 \times 2\farcs5$  & $8\farcs9 \times 8\farcs6$ & $3\farcs4 \times 3\farcs1$ & $10\farcs1~\times~ 9\farcs9$ & $35\farcs0 \times 32\farcs3$ \\
rms Stokes I~[$\rm\mu$Jy/beam]\tablefootmark{g} & 2.75  & 7.7  & 16 & 26  & 31\\
rms Stokes Q \& U~[$\rm\mu$Jy/beam]\tablefootmark{g} & 2.84 & 6.9  & - & 21 & 28\\ \hline
\end{tabularx}
\tablefoot{
\tablefoottext{a}{Bandwidth before flagging. At C-band the data cover 16 spectral windows (spws), i.e., 1024 spectral channels. At L-band, the data cover 32 spws (2048 spectral channels).}
\tablefoottext{b}{After Hanning smoothing.}
\tablefoottext{c}{Pointing 1: $\alpha_{2000}= 12^\mathrm{h}42^\mathrm{m}16.88^\mathrm{s}$, $\delta_{2000}~=~32^{\circ}32'37 \farcs2$. Pointing 2: $\alpha_{2000}= 12^\mathrm{h}41^\mathrm{m}59.14^\mathrm{s}$, $\delta_{2000}=32^{\circ}32'21\farcs6$.}
\tablefoottext{d}{Pointing: $\alpha_{2000}= 12^\mathrm{h}42^\mathrm{m}08.01^\mathrm{s}$, $\delta_{2000}=32^{\circ}32'29\farcs4$.}
\tablefoottext{e}{On-source observing time before flagging. At C-band, values represent the total observing time for both pointings.}
\tablefoottext{f}{Image results with Briggs weighting (robust = 0).}
\tablefoottext{g}{All rms values were extracted from non-PB corrected images.
At C-band, the rms was estimated from the mosaicked images by averaging the noise in areas observed with only one pointing and also in areas where both pointings overlap.}
}
\end{minipage}
\end{table*}

The primary calibrator 3C286 was used to determine the bandpass and the absolute polarization angle. To calibrate the instrumental polarization leakage, one scan was carried out on the unpolarized source OQ208 (J1407+2827). For the observation at C-band in the D array configuration the secondary calibrator J1310+3220 was used as a polarization leakage calibrator due to its wide range in parallactic angle ($>60^{\circ}$).

Due to the relatively small size of the primary beam in the C-band observations, which has a full width at half maximum (FWHM) of $\approx 7\farcm5$ as compared to the angular extent of NGC~4631 of 15$\arcmin$, we performed observations with two pointings to cover the entire galaxy.
The pointings were placed on each side of the center along the disk of the galaxy separated by 3$\farcm$75. At L-band, the FWHM of the primary beam is about 30$'$, hence, one pointing was sufficient. Total intensity and polarization calibration, as well as imaging of the data, were performed using the Common Astronomy Software Applications (CASA; \citealt{2007CASA}) package as described in detail in \cite{Irwin2013} and \cite{Wiegert2015}.
 
The data were imaged using the MS-MFS algorithm \citep{Rau2011}. All datasets were self-calibrated once in both amplitude and phase, verifying that the model to be used for self-calibration did not include any artifacts or any extended negative components. The self-calibration table made with the Stokes I image was then applied to the Stokes Q and U data. For all images, the Briggs' robust weighting scheme \citep{Briggs1995} was used (robust = 0 for all Stokes I images). In order to emphasize broad-scale structures at lower resolution in the Stokes I images, we utilized the u,v taper option within the cleaning algorithm. By combining datasets in different array configurations we improved the u,v coverage, the signal-to-noise ratio, and the spatial dynamic range. At C-band, both pointings were self-calibrated and imaged separately. Consequently, the primary beam (PB) correction was applied on each of the corresponding Stokes I pointings and both images were combined as described in \cite{Wiegert2015} (Sect. 3.3.5). We refer to the companion paper \cite{Mora-Partiarroyo2019_B} for results and analysis concerning the polarization data (hereafter denoted as Paper II).

\subsection{Effelsberg data at 1.42~GHz and 4.85~GHz}
\label{Effelsberg_observations}

Observations of NGC~4631 with the 100-m Effelsberg telescope centered at 4.85~GHz (0.5~GHz bandwidth) were performed in 2014 with the dual-beam receiver (Project 114-13). A total of 20 coverages in all three Stokes parameters I, U, and Q were made for a $45'\times35'$ scanning area, centered on the dynamical center of the galaxy (see Table~\ref{NGC}). Observations with the Effelsberg telescope at 1.42~GHz (bandwidth =
0.13 GHz) were made with the single-horn receiver during excellent weather conditions in April 2002 (Project 43-02) centered on the dynamical center of NGC~4631. A total of 16 coverages were taken, scanned in two orthogonal directions to reduce scanning effects.

\begin{figure*}[]
\centering
\begin{subfigure}{}
    \includegraphics[trim = 40mm 170mm 60mm 15mm, clip, width=1\columnwidth]{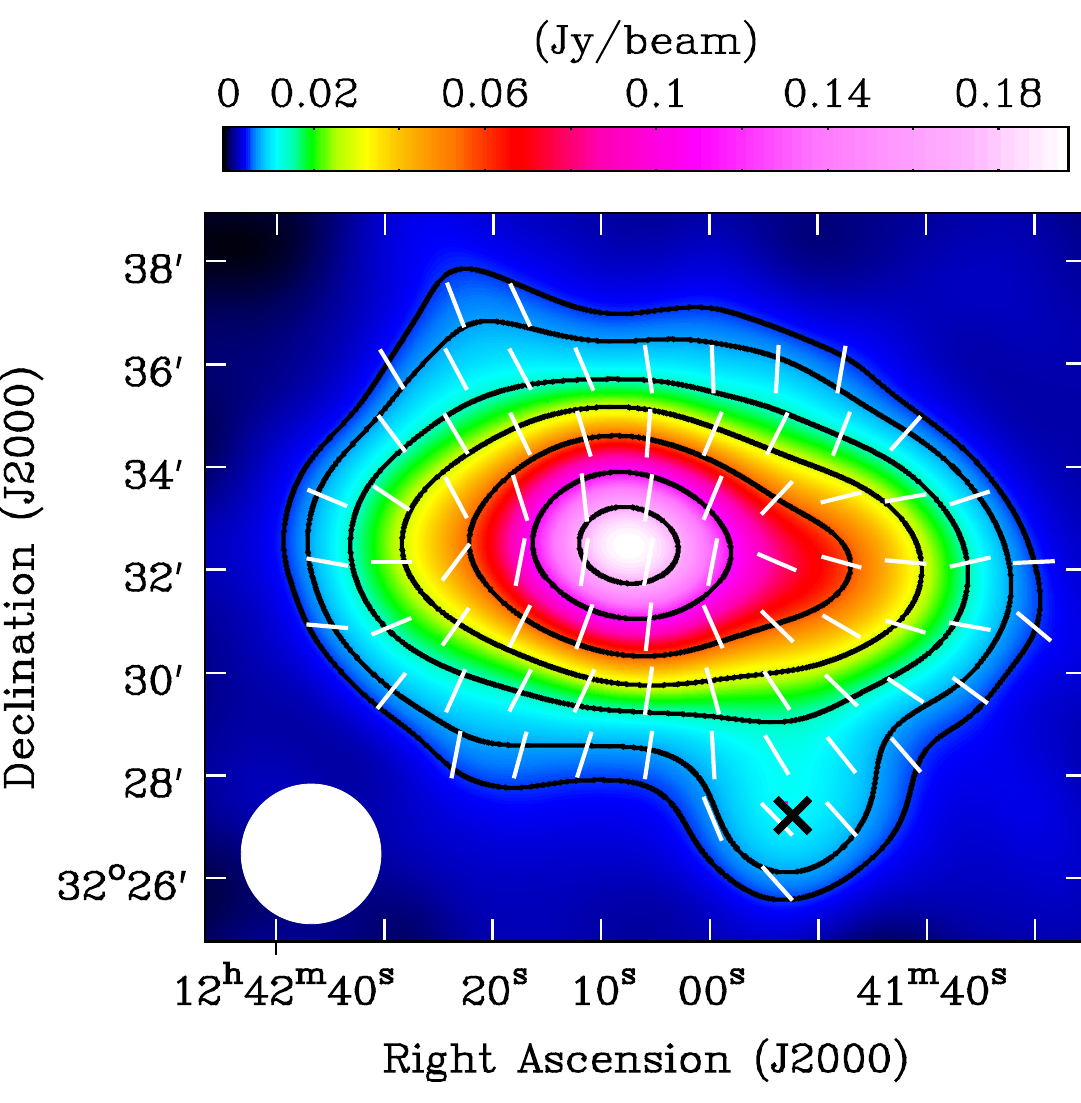}
\end{subfigure}
\begin{subfigure}{}
    \includegraphics[trim = 30mm 163mm 63mm 15mm, clip, width=1\columnwidth]{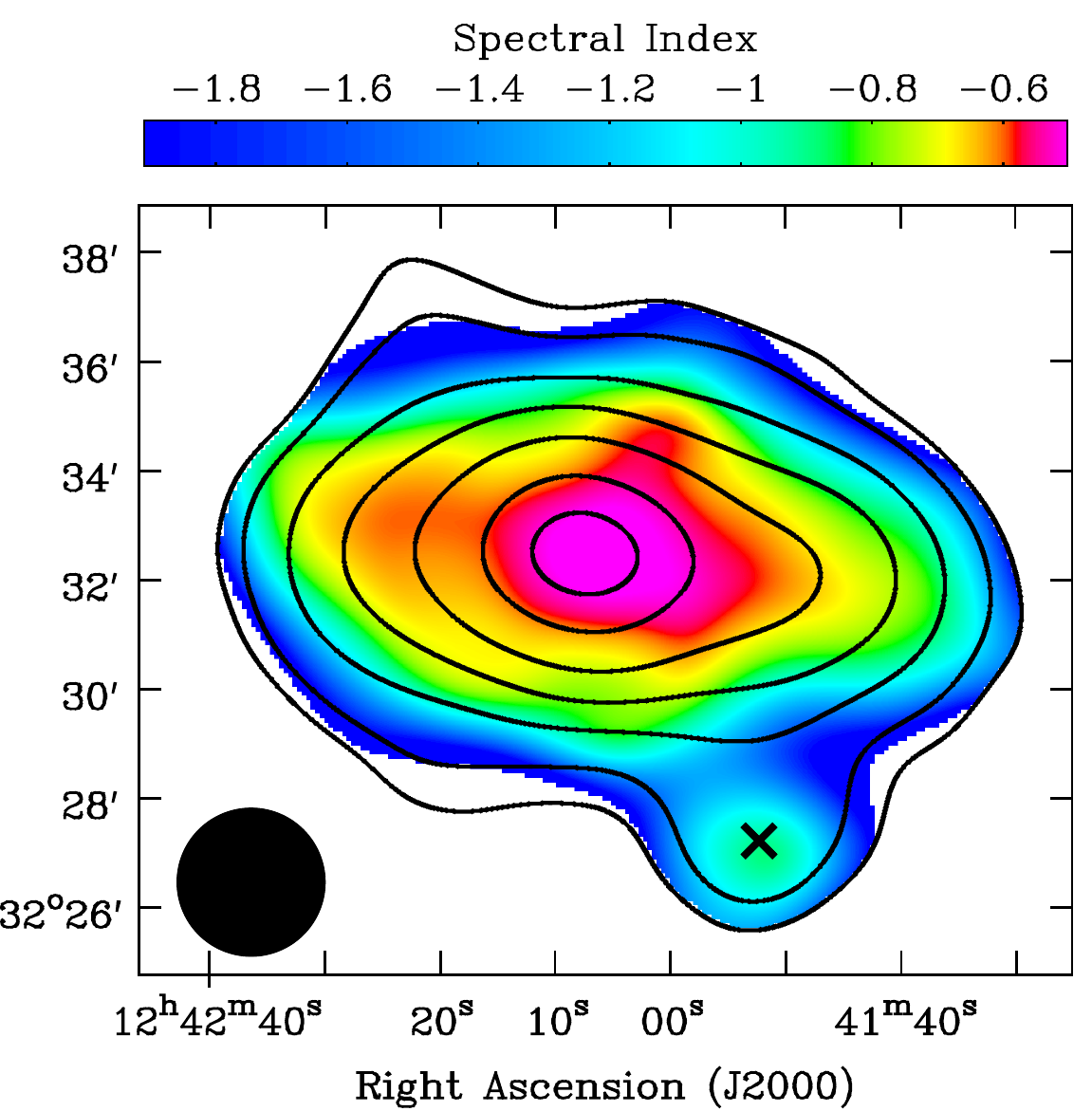}
\end{subfigure}
\caption[]{Left panel: Effelsberg total power emission of NGC~4631 at 4.85~GHz with apparent magnetic field orientation (line segments of equal length). Right panel: Spectral index distribution in NGC~4631 derived from Effelsberg data at 4.85 and 8.35~GHz (from \citealt{Mora2013}). All images have have an angular resolution of $163''$~FWHM. The contours correspond to the Effelsberg 4.85~GHz total intensity at $584~\rm{\mu Jy/beam}\times(8, 12, 24, 48, 96, 192, 288)$. The background radio galaxy to the southwest of NGC~4631 has been marked with a black X-symbol (see Appendix~\ref{BGRadioGalaxies}).}
\label{Effelsberg_observations_maps}
\end{figure*}

The Effelsberg data were reduced with the new NOD3 software package \citep{Mueller2017}, with which scanning effects due to weather conditions, receiver instabilities, and radio frequency interferences (RFIs) were removed. The scans of the galaxy at each observing frequency were combined and flux density calibration was done by using 3C\,286. We assume the 3C\,286 flux density scale as measured by the NRAO\footnote{\url{https://science.nrao.edu/facilities/vla/docs/manuals/oss/performance/fdscale}}. The zero-levels of the maps were corrected accordingly. At 4.85~GHz, the angular resolution of the map is $163''$~FWHM and the rms noise is $580~\rm{\mu Jy/beam}$ in total intensity and $140~\rm{\mu Jy/beam}$ in polarized intensity. The resulting total power emission is plotted in Figure~\ref{Effelsberg_observations_maps}. At 1.42~GHz the angular resolution is $561''$~FWHM and the rms noise is 33~mJy/beam in total intensity. At this resolution the galaxy is almost unresolved. 

\subsection{Integrated flux densities}
\label{Integrated flux densities}

In order to estimate the integrated flux densities, the intensities within the area enclosed by the 3$\sigma$ level were integrated in ellipses. Values are listed in Table~\ref{integrated_flux}. The uncertainties of the integrated flux densities include calibration and baselevel errors.

\begin{table}[h!]
\centering
\begin{minipage}{1\columnwidth}
\centering
\caption[Integrated flux densities of NGC~4631]{Integrated flux densities of NGC~4631. Values from this paper are marked in bold.}
\label{integrated_flux}
\begin{tabularx}{1\columnwidth}{c*3{>{\centering\let\\=\tabularnewline}X}}
\hline \hline
Telescope &  Frequency [GHz] & Integrated flux density [mJy]\\ \hline
Arecibo 300-m & 0.430 & 3072$\pm$95\tablefootmark{a} \\
Arecibo 300-m & 0.835 & 2005$\pm$80\tablefootmark{a} \\
\textbf{Effelsberg 100-m} & \textbf{1.42} & \textbf{1393$\pm$140}\tablefootmark{b} \\ 
\textbf{VLA D-configuration} & \textbf{1.57 (L-band)} &  \textbf{1091$\pm$40}  \\
Effelsberg 100-m & 2.68 & 790$\pm$40\tablefootmark{c} \\
Green Bank 140-ft & 4.75 & 546$\pm$30\tablefootmark{d} \\
\textbf{Effelsberg 100-m} & \textbf{4.85} & \textbf{515$\pm$40}\tablefootmark{e} \\ 
\textbf{VLA D-configuration} & \textbf{5.99 (C-band)} & \textbf{284$\pm$11} \\
Effelsberg 100-m & 8.35 & 310$\pm$16\tablefootmark{f} \\
Effelsberg 100-m & 10.55 & $265\pm12$\tablefootmark{g} \\
OVRO 40-m & 10.70 & 255$\pm$15$^d$ \\
Effelsberg 100-m & 24.50 & $172\pm15$\tablefootmark{h} \\
\hline
\end{tabularx}
\tablefoot{
\tablefoottext{a}{from \cite{Israel1983}.}
\tablefoottext{b}{This value overestimates the flux density of the galaxy by roughly 50~mJy, since it includes the flux density of several point sources in the vicinity which we were able to distinguish in the interferometric data.}
\tablefoottext{c}{\cite{Werner1988}}
\tablefoottext{d}{from \cite{Israel1983}. This value was not taken into account to derive the spectrum of the integrated emission (Sect. \ref{sync_spectrum}), because it can be replaced by a value obtained with the Effelsberg telescope at higher angular resolution.}
\tablefoottext{e}{
The flux densities in the 4.85~GHz Effelsberg map published by \cite{Mora2013} were underestimated, because the smaller size of their scanned area made it difficult to estimate an accurate baselevel.}
\tablefoottext{f}{\cite{Mora2013}}
\tablefoottext{g}{\cite{Dumke1995}}
\tablefoottext{h}{\cite{Klein2018}}
}
\end{minipage}
\end{table}

The flux densities presented in Table~\ref{integrated_flux} indicate that the VLA data at both frequencies suffer from missing spacings even in the most compact D-configuration. Therefore, single-dish data need to be combined with the VLA data to recover the missing flux density attributed to the incomplete u,v coverage of the interferometer. The spectrum of the integrated emission is discussed in Sect.~\ref{sync_spectrum}.

\subsection{Merging VLA and Effelsberg total intensity images}
\label{merging_section}

\begin{figure*}[h]
\centering
\includegraphics[trim = 0mm 0mm 0mm 0mm, clip, width=0.7\textwidth]{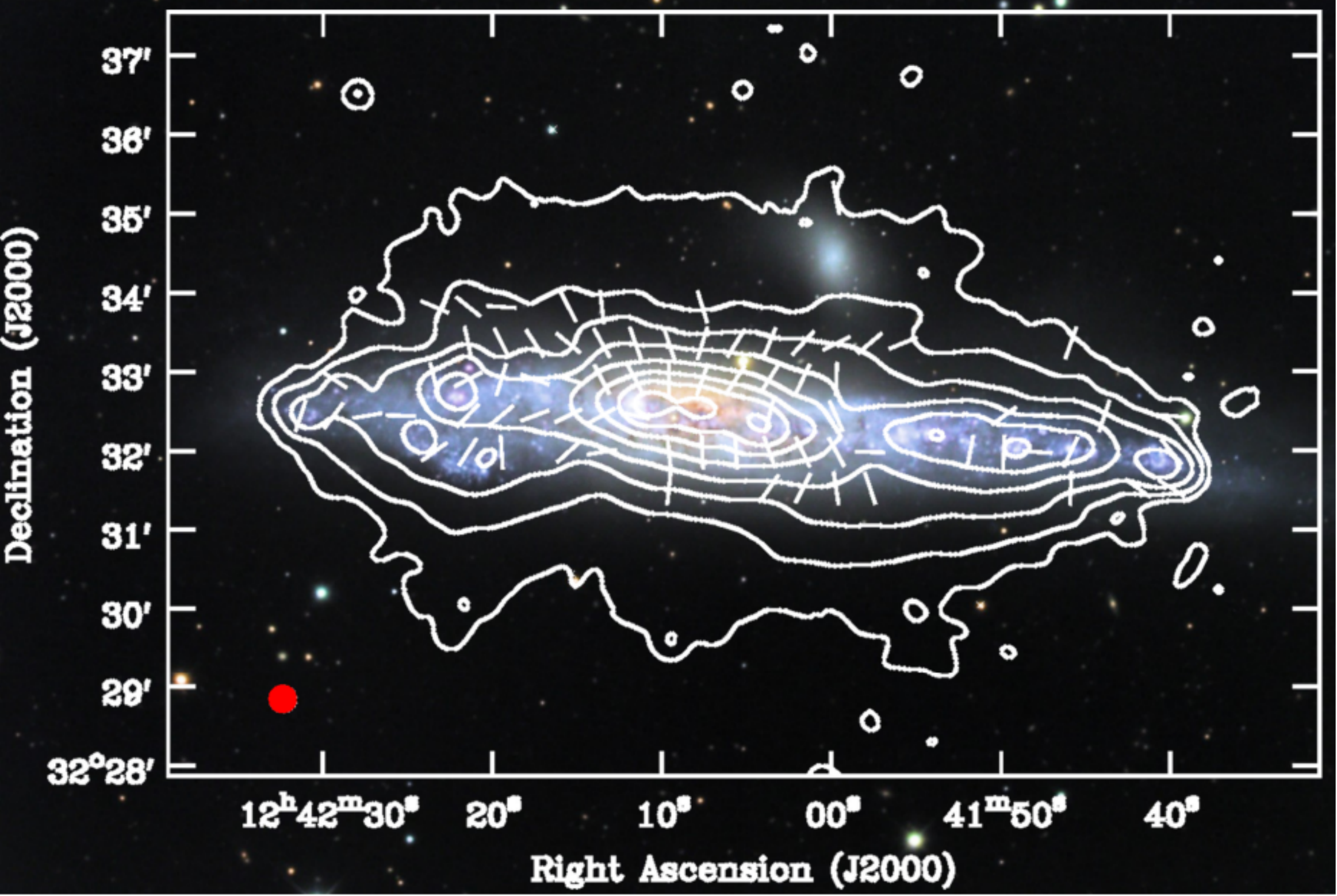}
\caption[Emission at 5.99~GHz (VLA + Effelsberg) over an optical Digitized Sky Survey image]{Total radio emission at 5.99~GHz (VLA + Effelsberg), overlayed on optical Digitized Sky Survey image (blue band). The angular resolution is $20\farcs5$~FWHM. The intrinsic magnetic field orientation (line segments of equal length) were obtained with RM-synthesis at pixels where the polarized intensity is larger than 5$\sigma$, where $\sigma$ is the rms noise in Stokes Q \& U (refer to Paper II).
Contour levels of the total power emission are at $45~\rm{\mu Jy/beam}\times(3, 6, 12, 24, 48, 96, 192, 384)$. The dwarf elliptical companion NGC~4627 can be observed in the northwestern quadrant of the galaxy. The triple radio source mentioned in Sect. 1, can be distinguished in the central region of the galaxy (eighth contour).}
\label{total_radio_over_optical}
\end{figure*}
 
\subsubsection{Rescaling of the Effelsberg data}

Prior to combining the images, the single-dish data were rescaled to the central frequency of the interferometric data. The Effelsberg data at 4.85~GHz were rescaled to 5.99~GHz by using the spectral index distribution between Effelsberg 4.85 and 8.35~GHz \citep{Mora2013} data shown in Figure \ref{Effelsberg_observations_maps}. The galaxy has an integrated flux density of 430$\pm$26~mJy at 5.99~GHz, given by the Effelsberg rescaled map.

The Effelsberg data at 1.42~GHz were rescaled to 1.57~GHz assuming an integrated total spectral index of $\rm{\alpha_{tot}= -0.81}$ (refer to Fig. \ref{integrated_spix}). Attempts were made to fit a spectral index distribution between the 1.42~GHz data and the Effelsberg map at 4.85~GHz (smoothed to 561$''$). However, due to the poor resolution, no coherent structures could be resolved, so the integrated spectral index of the galaxy had to be used instead. NGC~4631 has an integrated flux density at 1.57~GHz of 1285$\pm$130~mJy, determined by the Effelsberg map rescaled to this frequency. This value includes the contribution of unresolved point-sources ($\sim$50~mJy; refer to Table~\ref{integrated_flux}).

\subsubsection{Merging strategy}
\label{merging_strategy}

The Effelsberg and the VLA total intensity maps were merged in the image plane with the new NOD3 algorithm "ImMerge" \citep{Mueller2017}. This algorithm combines both maps by first creating a map of the difference of the interferometric map (smoothed to the resolution of the single-dish beam) from the single-dish map. This map is then rescaled by the ratio of the two beams (VLA beam/Effelsberg beam) and, consequently, added to the interferometric map at its original resolution. By combining the Effelsberg and VLA maps we recovered about 34\% at C-band and 12\% at L-band of the total flux density of the galaxy, which is attributed to the large-scale structures. All of the total intensity images presented in Sect.~\ref{Results} were obtained by combining Effelsberg and VLA data with the NOD3 algorithm. In Table \ref{noise_resolution} we have summarized the rms values of the merged maps at all the various resolutions presented in this paper.

To compare the merging procedure, we also tested combining the single-dish and interferometric data with the Astronomical Image Processing System (AIPS) task "IMERG" and the CASA task "feather". We obtain a similar result with the feather and ImMerge tasks, while with IMERG the emission of the galaxy is more extended in the northeastern and southwestern quadrants giving NGC~4631 a more asymmetric appearance. Since most edge-on galaxies studied up to now reveal a rather symmetric distribution of the radio halo emission \citep{Wiegert2015}, we believe that the ImMerge NOD3 algorithm and the feather CASA task are more accurate in this respect for our NGC~4631 data.

\begin{table}[h]
\begin{minipage}{1\columnwidth}
\centering
\caption[]{\small{Rms values of resulting merged Stokes I images (VLA + Effelsberg) at different resolutions presented in this paper.}}
\label{noise_resolution}
\begin{tabularx}{1\columnwidth}{c*2{>{\centering\let\\=\tabularnewline}X}}
\hline \hline
Resolution & rms$_{\rm 5.99GHz}$ [$\rm \mu$Jy/beam] & rms$_{\rm 1.57GHz}$ [$\rm \mu$Jy/beam]\\ \hline
7$''$ & 18 & -\\
15$''$ & 30 & 60 \\
18$''$ & 35 & 65 \\  
20\farcs5 & 45 & 70\\
35$''$ & 110 & 140 \\
51$''$ & - & 200 \\
\hline
\end{tabularx}
\end{minipage}
\end{table}

\subsection{Archival infrared and H$\alpha$ data}
\label{archival_data}

Ancillary 24~$\mu \rm{m}$ data for NGC~4631 were obtained from the Spitzer Infrared Nearby Galaxies Survey (SINGS; \citealt{Kennicutt2003}). The data were observed in 2005 with the Multiband Imaging Photometer for Spitzer (MIPS; \citealt{Rieke2004}). SINGS H$\alpha$ data were also available for NGC~4631; observations were performed with the 2.1-m telescope at the Kitt Peak National Observatory (filter: KP1563) in 2002 \citep{2003SINGS}. The data contain [NII] line emission, which causes an uncertainty in the H$\alpha$ intensity of about 10\%.

Infrared data of NGC~4631 at 70 and 160~$\mu \rm{m}$ were available from the project ``Key Insights on Nearby Galaxies: a Far-Infrared Survey with the Herschel'' (KINGFISH; \citealt{Kennicutt2011}). The source was observed with the \textsl{Photodetector Array Camera and Spectrometer (PACS; \citealt{Poglitsch2010})} on the {\it Herschel} space telescope. The 24, 70, and 160~$\mu \rm{m}$ data were convolved to a Gaussian beam using the kernels of \cite{Aniano2011}. All maps were transformed to the same coordinate grid and smoothed to a common resolution (18$''$ and 20\farcs5~FWHM) in order to derive the thermal emission of the galaxy presented in Sect.~\ref{thermal_nonthermal_separation}.

\section{Results}
\label{Results}


\subsection{Total power radio continuum: comparison with other wavebands}
\label{comparison_with_other_wavebands}

\begin{figure}[h]
\centering
\includegraphics[trim = 20mm 155mm 25mm 25mm, clip, width=1\columnwidth]{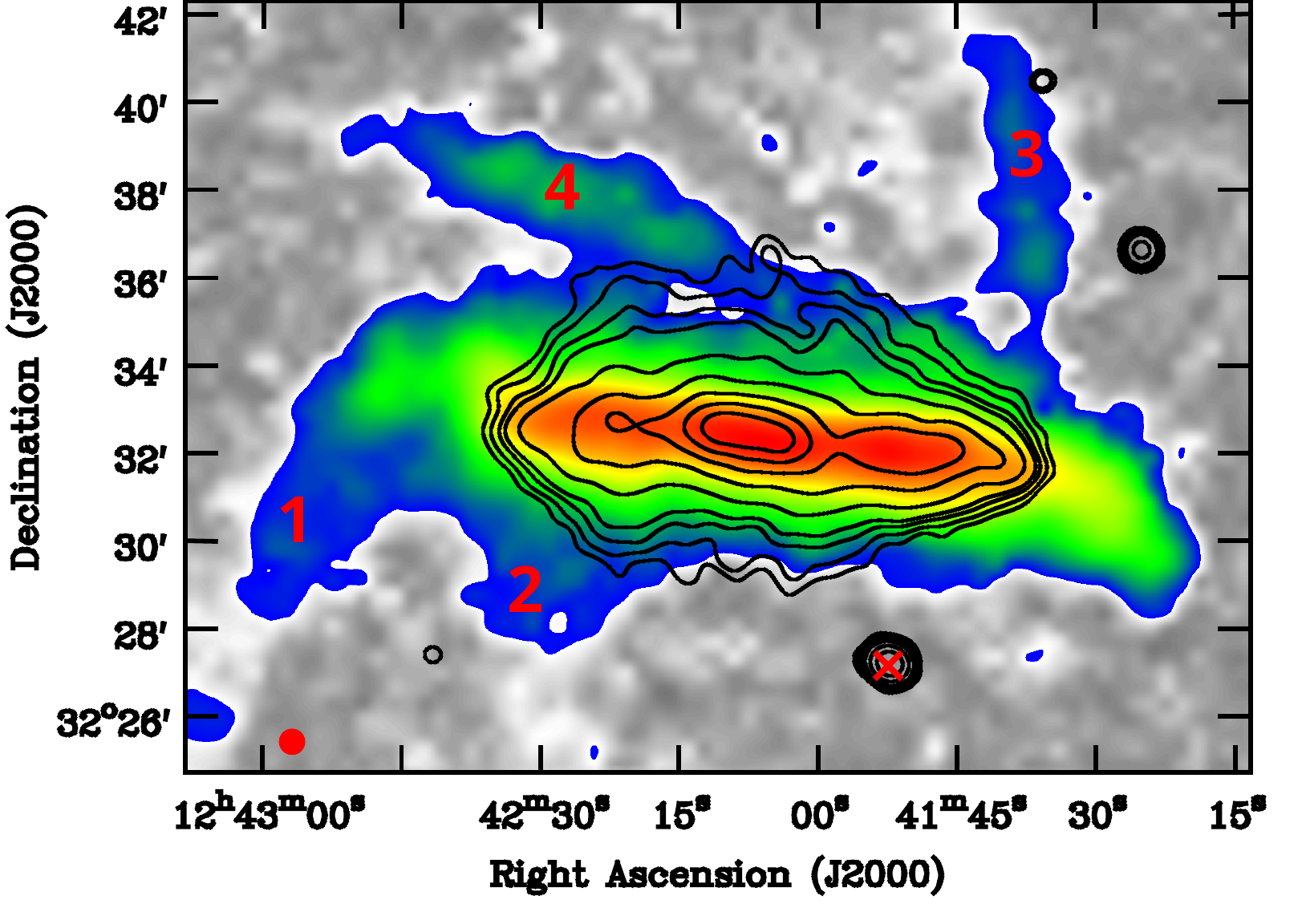}
\caption[Emission at 1.57~GHz (VLA + Eff.) over the HI radiation]{Total radio emission at 1.57~GHz (VLA + Effelsberg) overlaid on colorscale representation of integrated H{\sc i} emission from \cite{Rand1994}. All data plotted have an angular resolution of 35$''$~FWHM. Contour levels of the 1.57~GHz emission are at $140~\rm{\mu Jy/beam}\times(6, 8, 11, 14, 24, 48, 96, 192, 384)$. The H{\sc i} spurs have been labeled 1-4 following \cite{Rand1994}. The background radio galaxy to the southwest of NGC~4631 has been marked with a red X-symbol (see Appendix~\ref{BGRadioGalaxies}).}
\label{total_radio_over_HI}
\end{figure}

\begin{figure}[h]
\centering
\includegraphics[trim = 20mm 140mm 45mm 16mm, clip,width=\columnwidth]{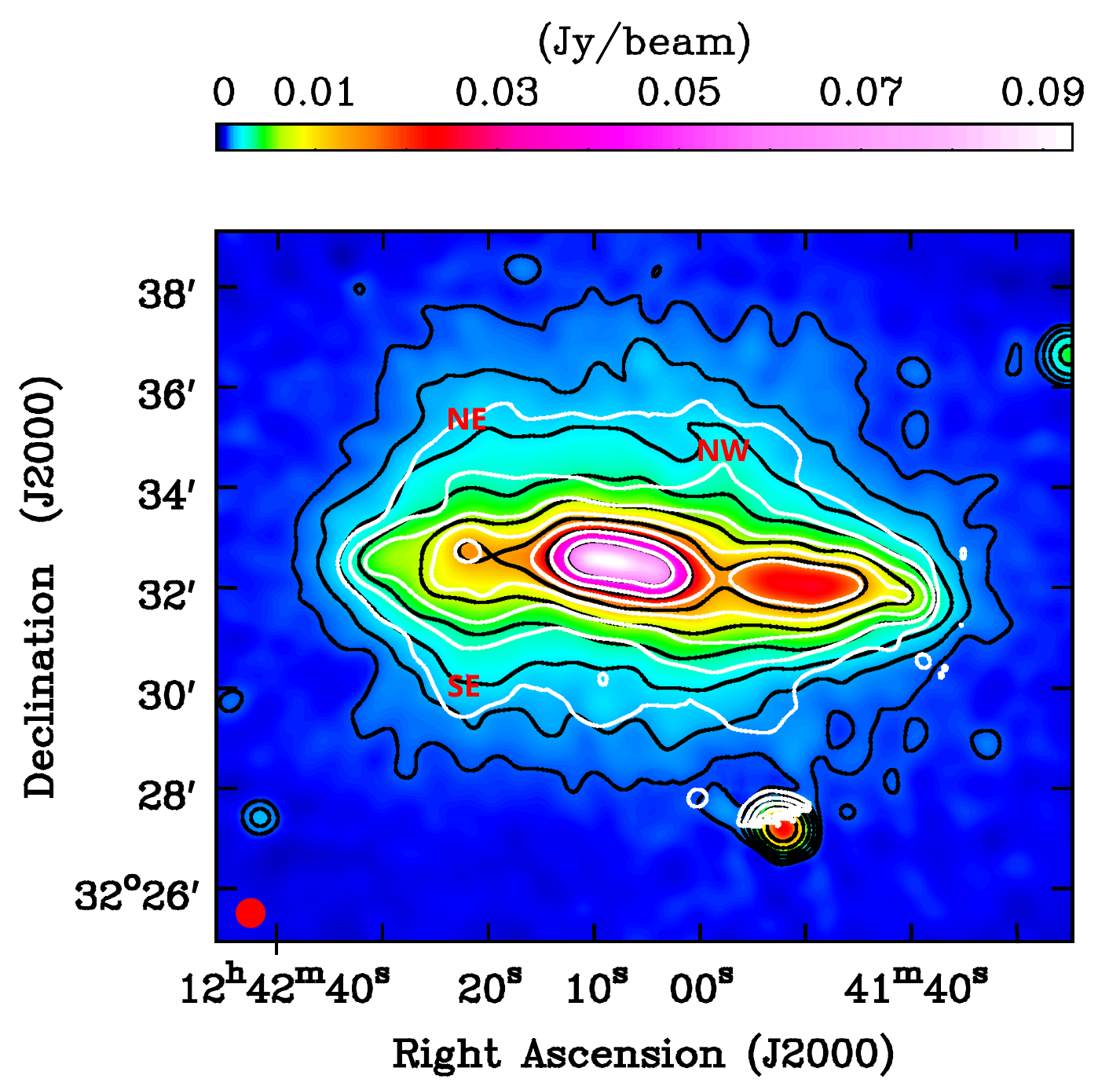}
\caption{Radio spurs observed at frequency of 1.57 and 5.99~GHz. The colorscale represents the 1.57~GHz emission (VLA + Effelsberg). White contours correspond to the 5.99~GHz emission (VLA + Effelsberg) are at $110~\rm{\mu Jy/beam}\times(3, 6, 12, 24, 48, 96, 192, 384)$. Black contours correspond to the 1.57~GHz radiation (VLA + Effelsberg) are at $140~\rm{\mu Jy/beam} \times(3, 6, 12, 24, 48, 96, 192, 384)$. All data plotted have an angular resolution of $35''$ FWHM. The SE-radio spur (SE), NE-radio spur (NE), and NW-radio spur (NW) have been indicated in the map. The total power map at 5.99~GHz is truncated in the southern part of the galaxy due to the primary beam. }
\label{total_radio_34arcs_C+Lband}
\end{figure}

Total intensity contours at 5.99 GHz of the galaxy are shown at a resolution of $20\farcs5$ in Figure~\ref{total_radio_over_optical} overlayed on an optical Digitized Sky Survey image, and at 1.57~GHz at a resolution of 35$''$ in Figure~\ref{total_radio_over_HI} together with the H{\sc i} emission \citep{Rand1994}. Contours of the total power emission at 5.99~GHz are displayed over the 1.57~GHz emission of NGC 4631 in Figure~\ref{total_radio_34arcs_C+Lband}.

 As can be seen in Figure~\ref{total_radio_over_optical}, the halo emission of NGC~4631 at 5.99~GHz has a symmetric distribution with respect to its minor axis, together with spur features in its southern halo. The distribution of the total intensity emission is considerably more symmetric with respect to the minor axis than in Fig.~4 of \cite{Mora2013} (which is at 4.85~GHz frequency and $\sim12\farcs5$ resolution). This could originate from the fact that in \cite{Mora2013} the data were combined by using the AIPS algorithm IMERG and in addition, due to insufficient baselevel determination in the Effelsberg data (as a result of the limited size of the scanned area), which resulted in missing flux densities, especially in the northwestern area of the galaxy.

The southeastern radio spur (hereafter denoted as "SE-radio spur"), identified by \cite{Hummel1990}, can be distinguished in Figures~\ref{total_radio_over_optical} and \ref{total_radio_34arcs_C+Lband} in much better detail. The prominent northwestern radio spur ("NW-radio spur"), which was discovered by \cite{Ekers1977}, is clearly distinguished in Figures~\ref{total_radio_over_HI} and \ref{total_radio_34arcs_C+Lband} at 1.57~GHz, and the northeastern radio spur (NE-radio spur) can also be observed in Figure~\ref{total_radio_over_HI}.
These two northern spurs seem to have counterparts at 5.99~GHz, but only at low resolution. \cite{Golla1994} speculated that the NW-radio spur originates close to the western peak of the triple radio source, but in our data (which are at higher resolution) this spur is clearly distinguishable only beyond $\sim$1\farcm5 above the plane ($\simeq$3.3~kpc). The NW-radio spur bends towards the northeast at more or less the projected position of the companion dwarf galaxy and it can be traced at 1.57~GHz up to about 3$\farcm$3 ($\simeq$7.4~kpc) above the center of the galaxy.

The H{\sc i} emission of NGC~4631 \citep{Rand1994} is shown in colorscale in Fig.~\ref{total_radio_over_HI} and the H{\sc i} spurs discovered by \cite{Weliachew1978} have been labeled 1--4. The projected length of the H{\sc i} spurs 1, 2, and 4 is about 60~kpc, while spur 3 reaches heights of about 30~kpc above the plane before bending back down (in velocity space) \citep{Rand1994}. The 1.57~GHz radio emission displays correlation with the H{\sc i} spurs 2 and 4. The SE-radio spur appears to be the radio counterpart of H{\sc i} spur 2. Furthermore, the uppermost emission of the NW-radio spur is oriented towards the northeast in the same manner as H{\sc i} spur 4. With respect to the dwarf elliptical companion NGC~4627 ($2\farcm6$ to the northwest of NGC~4631), there is radio emission that coincides with the position of the galaxy, however, no radiation can be unambiguously associated with this galaxy. \cite{Rand1994} came to the same conclusion for the H{\sc i} distribution.

From VLA test observations at 1.57~GHz (C-configuration, \citealt{Irwin2012}) the NW-radio spur is seen as a loop or partial loop which is just exterior to a loop-like feature seen in the X-ray emission in the northwestern halo of the galaxy (see Fig.~\ref{thermal_combination} of this paper and Fig.~7 of \citealt{Irwin2012}). Figure \ref{total_radio_over_HI} demonstrates the tight correlation between the uppermost emission of the NW spur and the H{\sc i} spur 4. According to \cite{Rand1994}, H{\sc i} spur 4 joins the disk at the western end and extends towards the northeast. This may indicate that the NW-radio spur might be composed of two features along the LOS that seem to look like one curved extension when projected onto the plane of the sky. One of the extensions may be oriented north--south having a counterpart at 5.99~GHz. The other extension oriented towards the northeast could be located in the outskirts of the halo closely connected to H{\sc i} spur 4. As pointed out by \cite{Irwin2012} the X-ray loop may be confined by the magnetic pressure of this radio spur.

\subsection{Thermal and nonthermal separation}
\label{thermal_nonthermal_separation}

To correct the H$\alpha$ radiation for extinction by dust we made use of a linear combination of the observed H$\alpha$ (L$_{\rm{H\alpha_{obs}}}$) and 24~$\rm{\mu m}$ (L$_{24\rm{\mu m}}$) luminosity from \cite{Kennicutt2003}:
\begin{equation}
\label{Halpha+24microns}
\rm{L_{H\alpha_{corr}}} = L_{H\alpha_{obs}} + a \cdot \nu_{24 \rm{\mu m}} \; \tilde{L}_{24 \rm{\mu m}},
\end{equation}
where the parameter $a$ is an unitless calibration factor. $L_{\rm H\alpha_{\rm corr}}$ and $L_{\rm H\alpha_{\rm obs}}$ are the corrected and observed H$\alpha$ line-integrated luminosities in units of $\rm erg\,s^{-1}$, while $\tilde{L}_{\rm 24\mu m}$
is the specific luminosity in units of $\rm erg\,s^{-1}\,Hz^{-1}$. \cite{Kennicutt2009} obtained a calibration factor of a~=~0.020$\pm 0.005$ from integrated measurements of galaxies, and \cite{Calzetti2007} obtained a value of a~=~0.031$\pm 0.006$ from H{\sc ii}-dominated regions. According to these authors, both values appear to be independent of metallicity. It has been suggested that the difference between the two calibration factors could be due to the primary stellar population heating the dust, that is, due to the IR cirrus problem \citep{Walterbos1996, Kennicutt1998}. In H{\sc ii} regions, the population of O-stars dominate gas ionization and dust heating. For entire galaxies, gas ionization is still dominated by O-stars, but the dust can also be heated by older stars. Therefore, in this paper, we employed the calibration factor a~=~0.031$\pm 0.006$ \citep{Calzetti2007}
for the luminosity range L$_{24 \rm{\mu m}} \geq 3\times10^{38}~\rm{erg\,{s}^{-1}}$ (typically for H{\sc ii} regions), while for lower luminosities the calibration factor a~=~0.020$\pm 0.005$ \citep{Kennicutt2009} was used.

We note, however, that in edge-on galaxies, mid-infrared emission at $24\,\mu$m could suffer from extinction. To account for this extinction, a slightly higher calibration factor (possibly dependent on column density)
is necessary \citep[see e.g.,][]{Vargas2018}. Employing a calibration factor of $a = 0.042$ \citep{Vargas2018} for a luminosity of L$_{24 \rm{\mu m}} \geq 3\times10^{38}~\rm{erg\,{s}^{-1}}$, would lead to a systematic increase of the overall integrated thermal flux density in NGC~4631 by only 16\% and is within the uncertainty of the method used in this paper (see below). 

The emission measure for an optically thick medium follows from the extinction-corrected H$\alpha$ emission \citep{Valls-Gabaud1998}. We adopt a typical value for the electron temperature of T$_\mathrm{e}$~$=10\,000$~K. Subsequently, the optical thickness and the brightness temperature of the thermal radio continuum emission at 1.57 and 5.99~GHz were derived as described in \cite{Fatemeh2007}. 

\begin{figure}[h]
\centering
\includegraphics[trim = 30mm 130mm 40mm 16mm, clip, width=1\columnwidth]{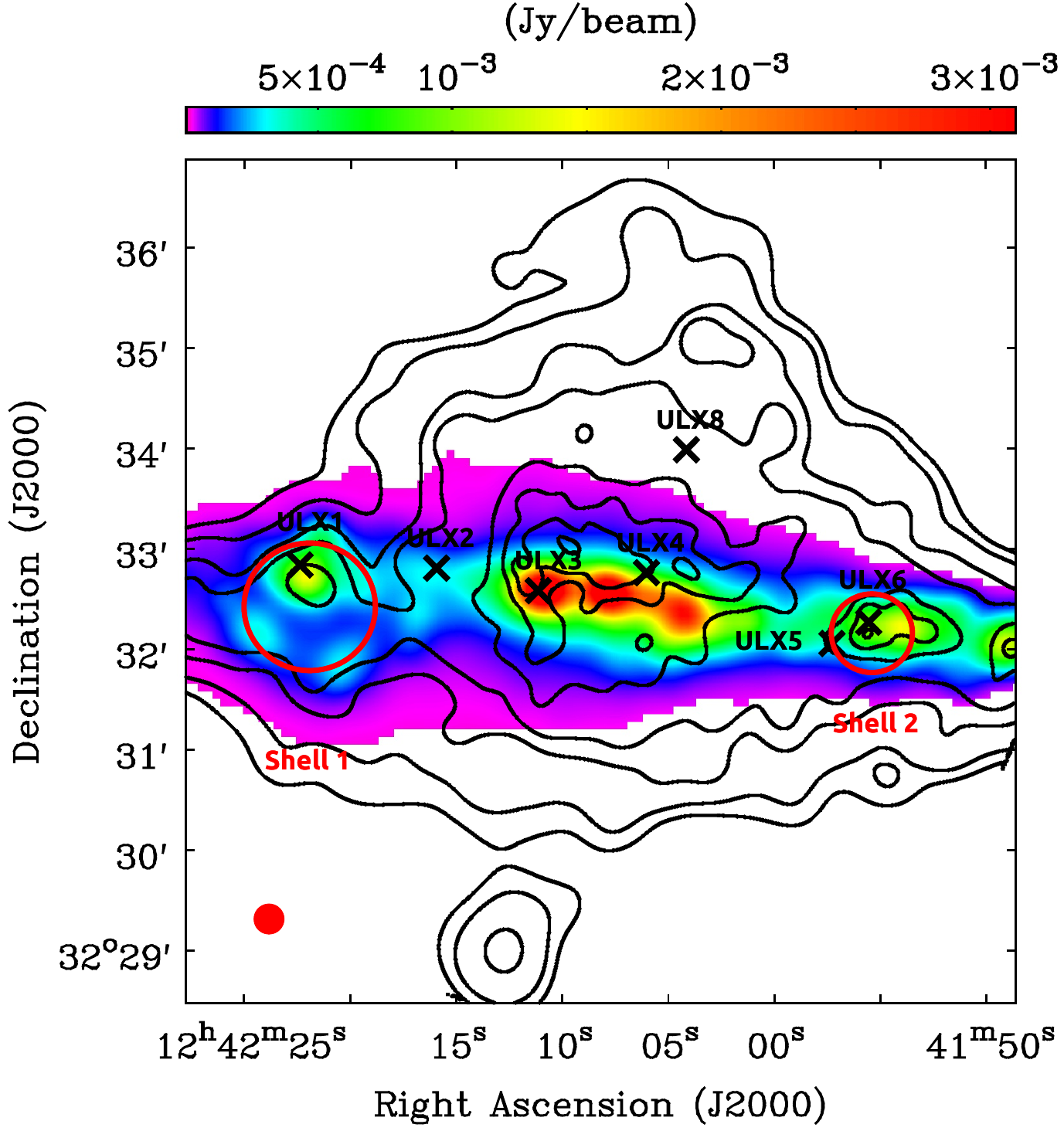}    
\caption[]{Colorscale of thermal emission at 5.99~GHz estimated from combination of H$\alpha$ and 24~$\mu$m data (18$''$~FWHM). Contours correspond to the soft X-ray emission (0.3--1.5~keV) from \cite{Wang2001}. The brightest X-ray sources \citep{Vogler1996, Read1997, Soria2009} are marked by X-symbols and the red circles represent the approximate size of the H{\sc i} supershells \citep{Rand1993}. }
\label{thermal_combination}
\end{figure}

The thermal emission of NGC~4631 is shown in Figure~\ref{thermal_combination} and the thermal fraction in Figure~\ref{thermal_fraction_M2}, at an angular resolution of $18\arcsec$~FWHM. Only pixels above a threshold of 3$\sigma$ in each of the maps were used. Locally within NGC~4631, the thermal contribution to the total radio emission is in the range 4--60\% (1--30\%) at 5.99 (1.57)~GHz. The highest values coincide with the H{\sc ii} regions CM~7, 8, and 9 \citep{Crillon1969}, located at the far western edge of the galaxy. Correction for extinction of the 24~$\mu$m data was not taken into account. As shown by \cite{Vargas2018}, the 24~$\mu$m extinction is predominant in NGC~4631 only in the central region of the galaxy. The estimated mean change in C-band thermal fraction from the inner disk to the outer disk correcting for extinction is about 14\% \citep{Vargas2018}.

The uncertainty in the thermal flux density is dominated by the adopted electron temperature (see \citealt{Fatemeh2007} for details). Considering an electron temperature variation of $\pm$2000~K and, in addition, taking into account the rms noise of the maps and the uncertainty introduced by the calibration factors, we obtain via error propagation that the thermal flux densities have an average relative error of about 19\% (17\%) at 5.99 (1.57)~GHz.

\begin{figure}[h]
\centering
\includegraphics[trim = 12mm 165mm 48mm 16mm, clip, width=1\columnwidth]{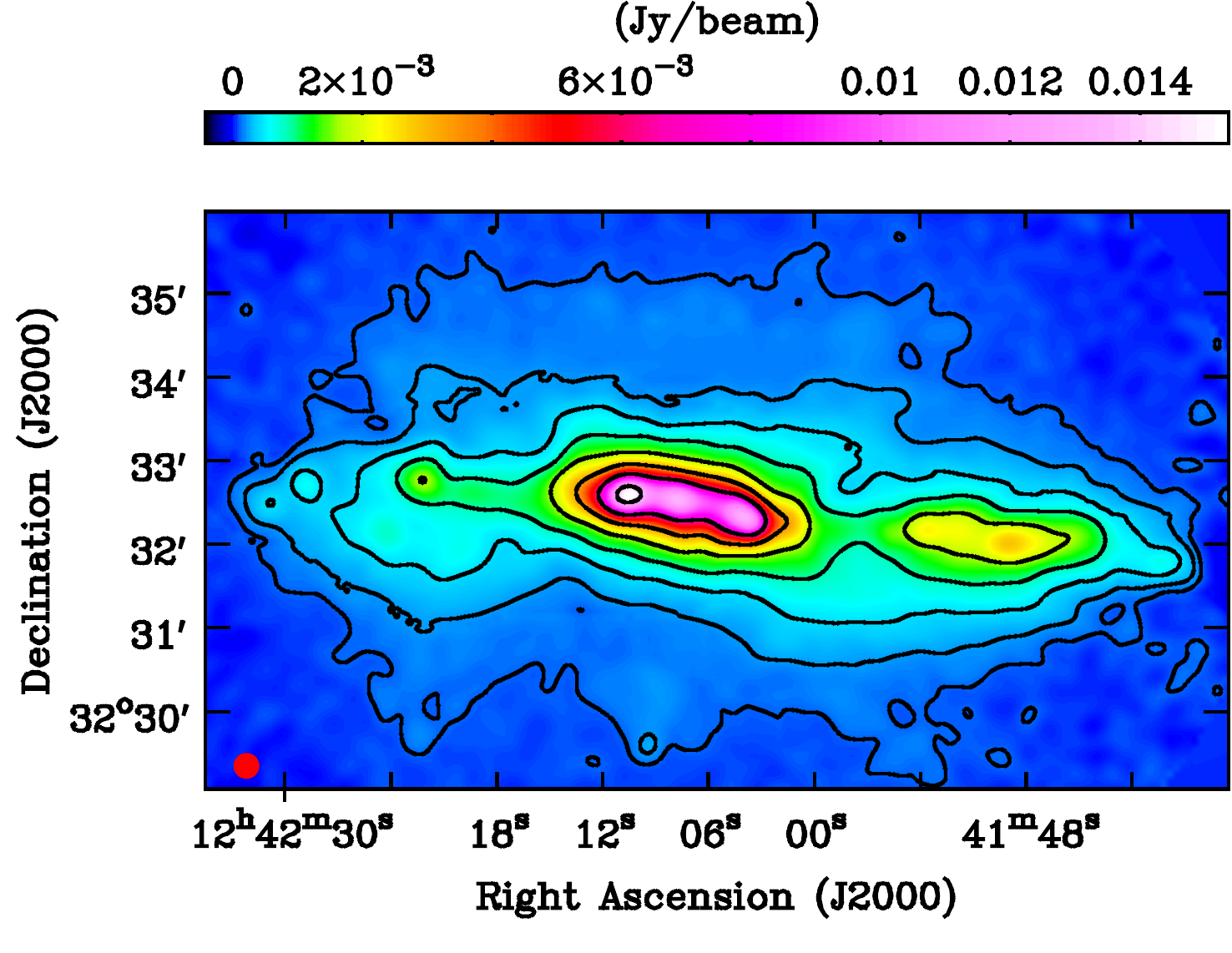}
\caption[]{Synchrotron emission of NGC~4631 at 5.99~GHz (VLA + Effelsberg) in colorscale and contours. The angular resolution is $18''$~FWHM. Contours are at $35~\rm{\mu Jy/beam}\times(3, 6, 12, 24, 48, 96, 192, 384)$.}
\label{synchrotron_emission_Cband}
\end{figure}

Integrating the thermal emission of NGC~4631 in ellipses, we obtain a galaxy-integrated thermal flux density of 61$\pm$13~mJy at 5.99~GHz and 70$\pm$13~mJy at 1.57~GHz. By integrating the emission of the entire galaxy we obtain a global  thermal fraction of 14$\pm$3\% at 5.99~GHz and 5.4$\pm$1.1\% at 1.57~GHz. These values agree with those published by \cite{Mora2013} if we interpolate their thermal fraction at 4.85~GHz (which was obtained by adding the radio and optical contributions). Fitting the flattening of the integrated spectrum due to thermal emission, \cite{Fatemeh2017} obtained a global thermal fraction of 10$\pm$4.5\% at 1.4~GHz (corresponding to 11$\pm$5\% at 1.57~GHz), which is consistent with our value at 1.57~GHz within the uncertainties. 

The synchrotron emission at 5.99~GHz obtained by subtracting the thermal emission from the total radio emission on a pixel-by-pixel basis is shown in Figure~\ref{synchrotron_emission_Cband} at an angular resolution of $18\arcsec$ FWHM. The synchrotron map at 1.57~GHz is not shown here because the thermal fraction is considerably small at this frequency. 

\begin{figure*}[]
\centering
  \begin{subfigure}{}
   \includegraphics[trim = 29mm 155mm 45mm 15mm, clip, width=1\columnwidth] {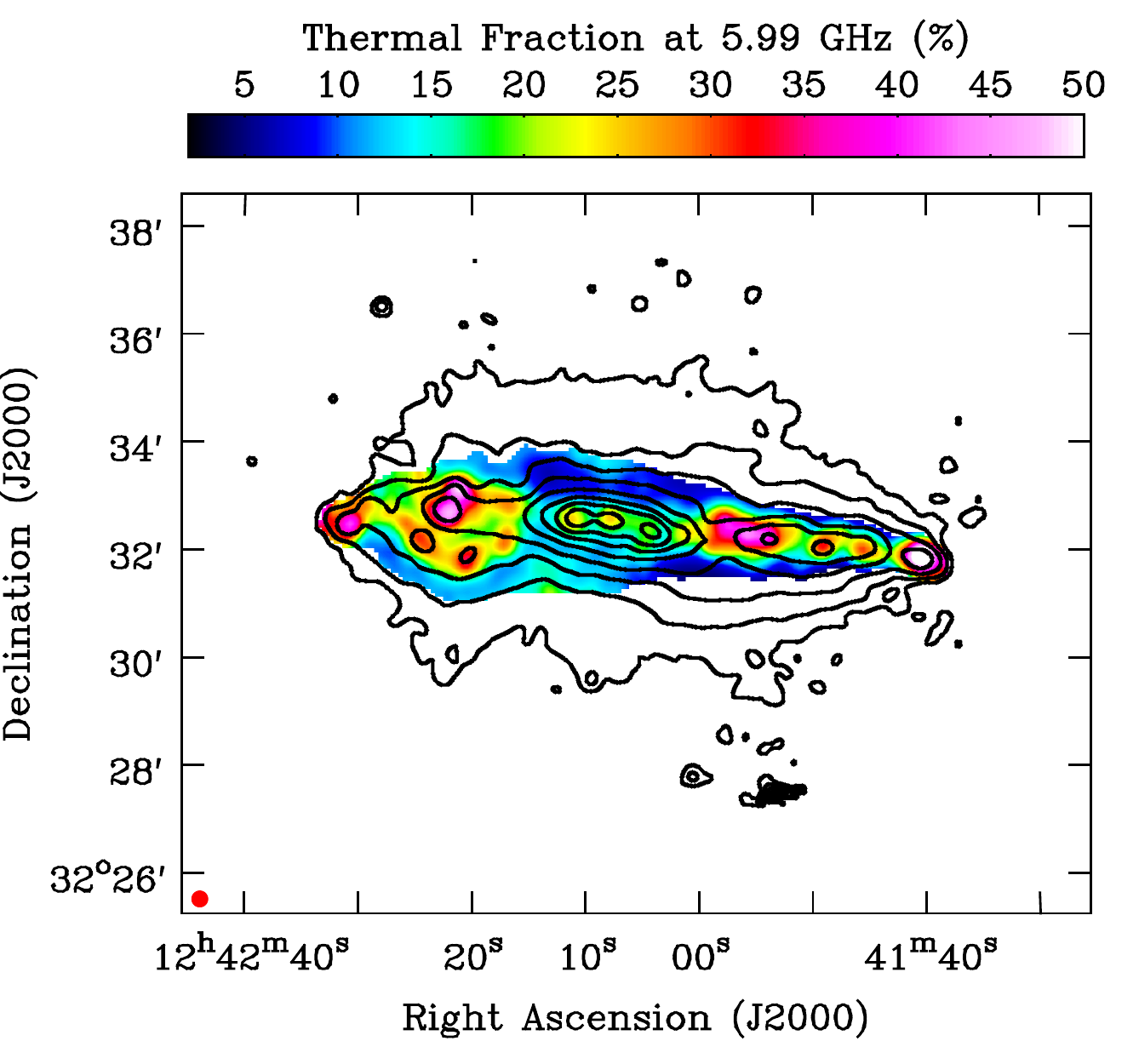}
  \label{Cband_method2_fraction}
  \end{subfigure}
  \begin{subfigure}{}
  \includegraphics[trim = 30mm 155mm 45mm 15mm, clip, width=1\columnwidth]{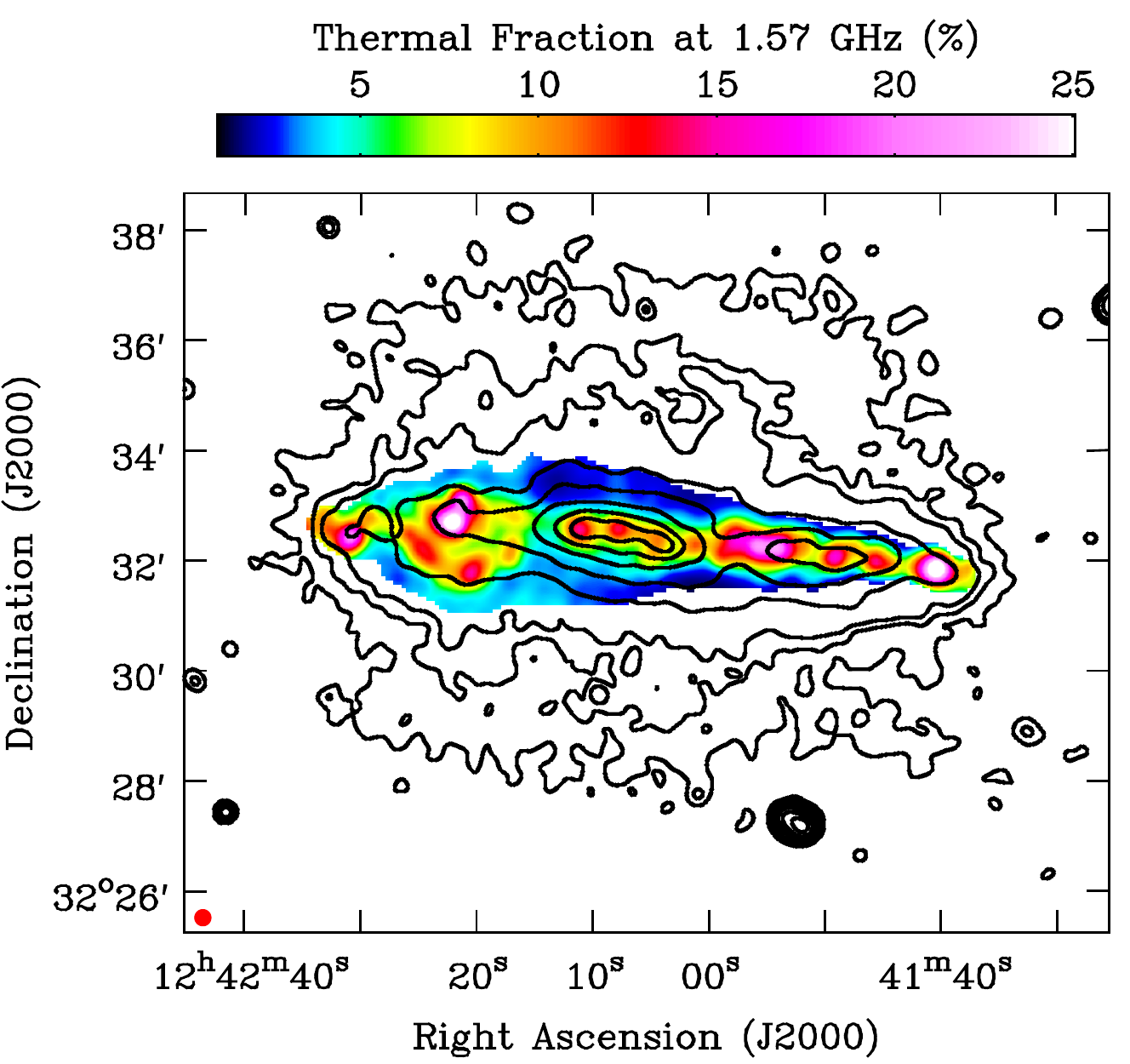}   
  \label{Lband_method2_fraction}
  \end{subfigure}
\caption[Thermal fraction (in color) estimated from combination of H$\alpha$ and 24~$\rm{\mu m}$ data (method two)]{Thermal fraction (in color) estimated from combination of H$\alpha$ and 24~$\rm{\mu m}$ data. The angular resolution is 18$''$~FWHM. Left panel: thermal fraction at 5.99~GHz with contours of the total 5.99~GHz emission (VLA + Effelsberg) at $35~\rm{\mu Jy/beam}\times(3, 6, 12, 24, 48, 96, 192,384)$. Right panel: thermal fraction at 1.57~GHz with contours of the total 1.57~GHz emission (VLA + Effelsberg) at $65~\rm{\mu Jy/beam}\times(3, 6, 9, 24, 48, 96, 192, 384)$.}
\label{thermal_fraction_M2}
\end{figure*}

\subsection{Spectrum of the integrated radio continuum emission}
\label{sync_spectrum}

To study the radio continuum spectrum of NGC~4631, we use only integrated flux densities obtained from single-dish telescopes (Table \ref{integrated_flux}). Extrapolation of the thermal flux density determined in Sect. \ref{thermal_nonthermal_separation} to 1.0~GHz yields 73$\pm$14~mJy and a thermal fraction of 4.0$\pm$0.9\%, which is smaller by a factor of 3.5$\pm$0.8 compared to the value of 14$\pm1$\% obtained by fitting the spectrum of total emission by a nonthermal component with a spectral cutoff plus a thermal component \citep{Klein2018}. This large discrepancy cannot be explained by the uncertainties of our method of measuring the thermal emission. We suspect that the quality of the spectral data of NGC~4631 is not sufficient to reliably measure the thermal fraction and the cutoff frequency independently from the radio spectrum alone.

\begin{figure}[h]
\centering
\includegraphics[trim = 5mm 5mm 10mm 20mm, clip, width=0.95\columnwidth]{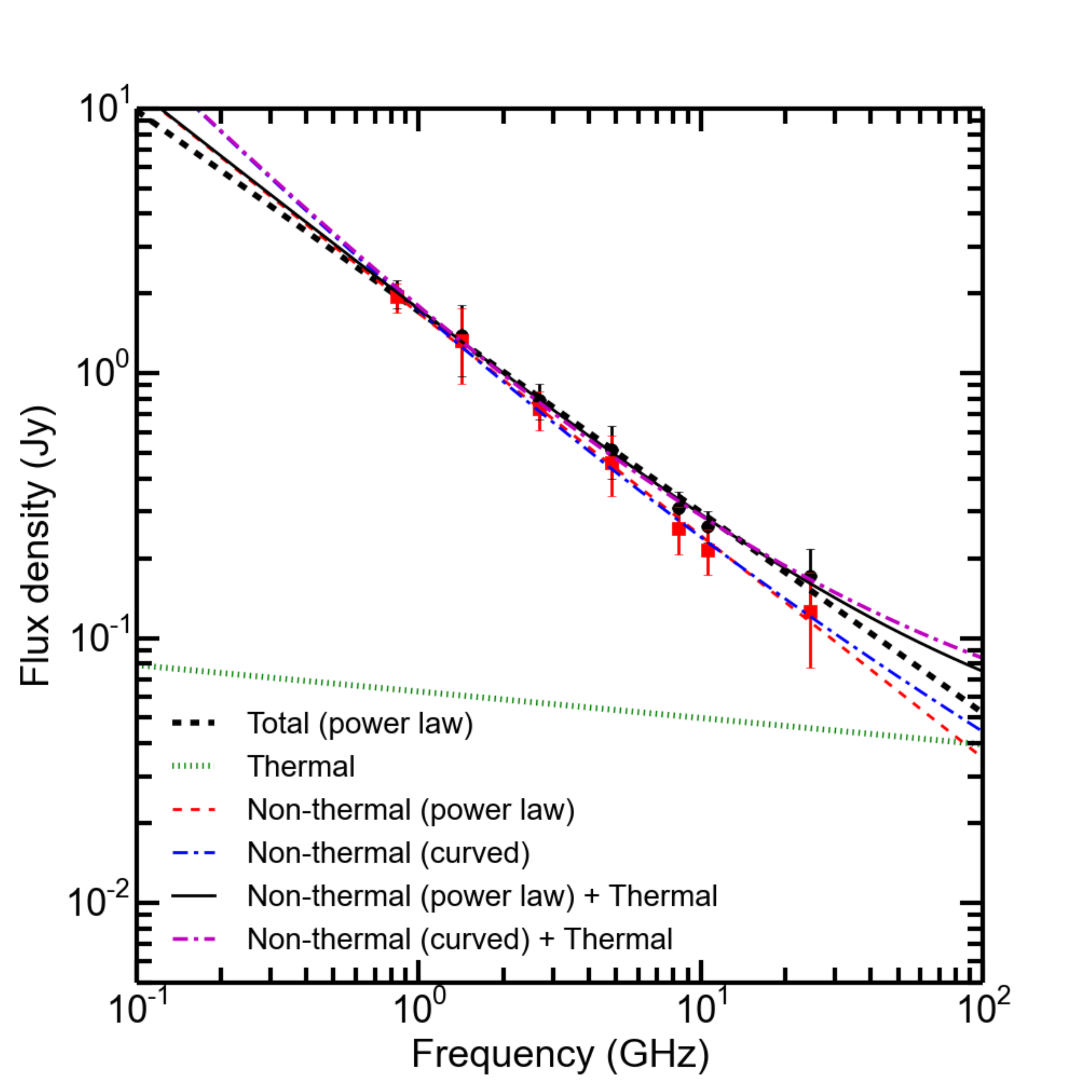}
\caption[Integrated total radio spectrum]{Integrated radio emission of NGC~4631: total (black), thermal (green), and synchrotron (red). The red dashed line shows the best-fit power law, with a synchrotron spectral index of $\alpha_\mathrm{syn}= -0.84 \pm 0.02$. Only single-dish data (Table~\ref{integrated_flux}) were taken into account for this calculation.}
\label{integrated_spix}
\end{figure}

We subtracted the thermal emission from the total flux densities obtained with single-dish data only, based on the thermal flux density of $S_\mathrm{th}=~73$~mJy at 1.0~GHz and the thermal spectral index of $-0.1$. By fitting a power law of the form $S_\mathrm{\nu,nth} \propto \nu^{\alpha_\mathrm{syn}}$ we estimated the spectral index of the nonthermal (synchrotron) emission of the galaxy. In Figure~\ref{integrated_spix}, the red dashed line represents the best fit power law, with a synchrotron spectral index of $\alpha_\mathrm{syn}= -0.84 \pm 0.02$. We emphasize that there is no indication for a break or cutoff in the synchrotron spectrum, as argued by \cite{Klein2018}. The cutoff in the synchrotron spectra found by \citet{Klein2018} is perhaps due to their highly overestimated thermal emission.

The energy of the spectral break at the transition from dominating bremsstrahlung loss to dominating synchrotron loss depends on gas density and magnetic field strength. As both quantities vary over the galaxy considerably, the energy of the spectral break also varies, so that any break would be washed out in the radio spectrum \citep{Basu2015}.

\subsection{Thermal emission: comparison with X-ray data}

In Figure~\ref{thermal_combination}, the estimated thermal emission map of NGC~4631 at 5.99~GHz is shown as a colorscale together with contours of the soft X-ray emission \citep{Wang2001}. The brightest X-ray emitting feature is located above the central region. \cite{Wang1995} found that this bright X-ray feature appears to be connected to H$\alpha$ filaments emerging from the galaxy's disk. The red circles in Figure~\ref{thermal_combination} indicate the approximate size of the H{\sc i} supershells found by \cite{Rand1993}. Shell 2 (towards the west of CM~67), as labeled in \cite{Rand1993}, seems to be an expanding bubble, whereas Shell 1 (towards the east of CM~67) is probably caused by an impact of a cloud \citep{Neininger1999}. Shell 2 corresponds to the distortion in the eastern side of the galactic disk appearing in all optical, far-ultraviolet (FUV), and H$\alpha$ maps \citep{Smith2001}.

The bright thermal feature ($\alpha_{2000}=12^\mathrm{h}42^\mathrm{m}22^\mathrm{s}, \delta_{2000}=32^{\circ}32'50 \farcs5$) to the east of the central region, where the H{\sc ii} region CM~81 is located, coincides not only with an X-ray and radio peak, but also with the H{\sc i} supershell 1. Similarly, the H{\sc i} supershell 2, located close to the H{\sc ii} regions CM~37, 38, roughly coincides with an emission peak seen in the thermal, X-ray, and radio ($\alpha_{2000}= 12^\mathrm{h}41^\mathrm{m}55^\mathrm{s}, \delta_{2000}=32^{\circ}32'16 \farcs5$).

\begin{figure*}[h!]
\centering
\includegraphics[width=1.9\columnwidth]{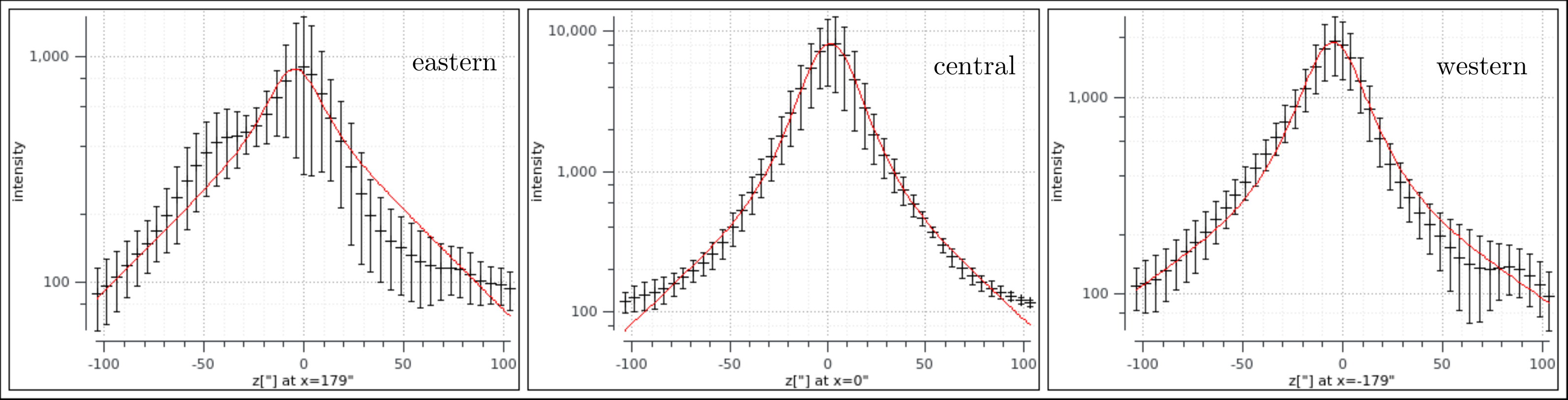}
\caption[]{Averaged intensity distributions of total emission at 5.99~GHz (VLA + Effelsberg) in boxes along three vertical strips of 180$''$ ($\simeq6.6$~kpc) width, together with best two-component exponential fits (red lines).}
\label{scale height}
\end{figure*}

The extraplanar X-ray emission in Fig.~\ref{thermal_combination} could be related either to regions in the underlying galactic disk with strong thermal emission (high thermal pressure) or to regions with strong synchrotron emission (Fig.~\ref{synchrotron_emission_Cband}) (high nonthermal pressure). Intense starburst regions responsible for the extraplanar X-ray emission have stronger emission in every band, and a larger fraction of the energy may go into cosmic rays driving the radio halo.

\subsection{Vertical scale heights}
\label{section_scale_heights}

We determined the vertical radio scale heights of NGC~4631 with the NOD3 package from emission profiles that were obtained by averaging the intensity along strips perpendicular to the major axis of the galaxy. These profiles were fit with a model distribution consisting of a function with two exponentials convolved with the effective beam. The effective beam is composed of the FWHM of the observations and a correction for the disk emission for galaxies that are seen with inclinations $i < 90\degr$, as described in \cite{Mueller2017} and \cite{krause+2018}.

The vertical scale heights were determined at 1.57 and 5.99~GHz with the combined maps (VLA + Effelsberg) at an angular resolution of 15$''$ in total emission and 18$''$ in synchrotron intensity. Inclinations of NGC~4631 between $84^{\circ}$ or $87^{\circ}$ are given in the literature \citep{Vaucouleurs1963,Rand1994}. From our fits to the vertical intensity distributions we could clearly see that NGC~4631 is more inclined. We could get reasonable fits to our data only if we assume the inclination of NGC~4631 to be $ \ge 88^{\circ}$. In fact, Irwin et al. (2011) modeled the central molecular ring in NGC~4631 with a best-fitting inclination of $89^{\circ} \pm 4^{\circ}$. From our vertical profiles of NGC~4631, we conclude an inclination of $89^{\circ} \pm 1^{\circ}$ for the entire galaxy. It was noted already by \cite{krause+2018} that the fits to the averaged vertical intensity distribution are very sensitive to the value of the inclination in other galaxies of the CHANG-ES sample.

The intensity averaging of the NGC~4631 maps was estimated along three strips parallel to the minor axis with a width of 180$''$ each, centered on the nucleus. This relatively large strip width allows averaging over small-scale asymmetries known in NGC~4631, and allows the analysis of the radio spurs in the eastern and western side of the galaxy separately. With the NOD3 program, we fit the intensity profiles above and below the disk of NGC~4631 simultaneously. A two-component exponential function fit the data best at both wavelengths, instead of a two-component Gaussian function or one-component fits. The corresponding values for the scale heights for the thin disk and halo for the total- and synchrotron-intensity profiles are summarized in Table~\ref{scale_heights} and Table~\ref{scale_heights_synchrotron}, respectively. 

As an example, we show the vertical intensity distributions together with the best fits for the total intensity at 5.99~GHz in Fig.~\ref{scale height}. The central intensity distribution looks very symmetric above ($z > 0$) and below ($z < 0$) the major axis with the maximum being centered on the major axis ($z=0$), while the eastern (left) and western (right) distributions have local asymmetries. The peak of the observed distribution in the western strip is shifted by about $10 \arcsec$ south of $z=0$, reflecting the warp of the disk of NGC~4631. The eastern strip displays the strongest asymmetries and hence, has the largest deviations from the two-component exponential fit: there is an excess of radio emission below the major axis and a depletion above the midplane. This causes the peak of the fit curve to shift to the south, although the maximum of the observed averaged intensity distribution is at $z = 0$ and therefore, it does not indicate a warp of the disk at the eastern side.

At 5.99~GHz, the total power emission of NGC~4631 is characterized by a thin disk with a mean scale height ($\rm h_{thin}$) of $0.35\pm0.62$~kpc, while for the synchrotron emitting thin disk the mean scale height is $0.37\pm0.34$~kpc. In the halo, we find a mean scale height ($\rm h_{halo}$) of $1.8\pm0.5$~kpc and $1.6\pm 0.4$~kpc for the total and synchrotron emission, respectively. As the scale heights for the synchrotron emission at 5.99~GHz in the eastern strip (E in Table~\ref{scale_heights_synchrotron}) strongly deviate and have extremely large errors, we omitted them when calculating the mean values. 
At 1.57 GHz, the thin disk of NGC\,4631 has a mean scale height of $0.39\pm0.02$ and $0.34\pm0.08$ kpc for the total and synchrotron intensity, respectively. The halo of NGC\,4631 at 1.57 GHz has total intensity scale height of $1.8\pm0.3$ kpc and synchrotron scale height of $1.7\pm0.3$ kpc.

The total intensity scale heights obtained at 5.99~GHz for the different strips within NGC~4631 differ from the values found by \cite{Mora2013}. This is because the emission of the galaxy has a more asymmetric distribution in the 4.85~GHz combined map published in \cite{Mora2013}. Especially in the halo, the mean scale heights (h$_{\rm{thin}} = 0.4\pm 0.3$~kpc and h$_{\rm{halo}} = 2.3 \pm 0.9$~kpc) published in \cite{Mora2013} are larger. Nevertheless, the values agree within the errors with our mean scale heights obtained at 5.99~GHz in total power.

The largest deviations from a  symmetrically averaged intensity distribution are found in the eastern strip at distances larger than $20''$ from the midplane (see Fig.~\ref{scale height}). The excess in the southeast might be related to the higher star formation activity in this region with respect to the other areas of the galaxy. The distribution of thermal emission (see Figs.~\ref{thermal_combination} and \ref{thermal_fraction_M2}) indicates that in the eastern side of the disk there is significant emission extending south of the midplane. This excess of the averaged radio intensity below the disk together with the depletion north of the midplane, causes the peak of the scale height fit to be shifted about $10''$ south of the midplane as mentioned above.

The asymmetries are also visible in the synchrotron intensity in the eastern strip where the excess in the south is even more smeared out. This is probably not due to the slightly lower resolution of the synchrotron maps, but it is expected since the total radio emission is influenced by thermal emission and the cosmic-ray electrons (CREs) that produce synchrotron emission propagate farther away from star-forming regions than the non-relativistic electrons that produce thermal emission.  Within the errors, however, we do not find that the mean synchrotron scale heights are in general larger than the mean scale heights of the total radio emission.
 
\begin{table}[h]
\centering
\begin{minipage}{1\columnwidth}
\caption[]{Vertical scale heights of thin ($\rm{h_{thin}}$) and halo ($\rm{h_{halo}}$) components of exponential fits to total intensity profiles. The eastern (E), central (C) and western (W) strips have a width of 180$''$ each.}
\label{scale_heights}
\begin{tabularx}{1\columnwidth}{c*5{>{\centering\let\\=\tabularnewline}X}}
\hline \hline
\multirow{2}{*}{Strip} & \multicolumn{2}{c}{5.99~GHz} & \multicolumn{2}{c}{1.57~GHz} \\
                                &  $\rm{\bm h_{\bm t\bm h\bm i\bm n}}$ [pc] & $\rm{\bm h_{\bm h\bm a\bm l\bm o}}$ [kpc] & $\rm{\bm h_{\bm t\bm h\bm i\bm n}}$ [pc] & $\rm{\bm h_{\bm h\bm a\bm l\bm o}}$ [kpc]\\ \hline
E &   316$\pm$247   & 1.82$\pm$0.43 & 397$\pm$528   &  2.05$\pm$0.31  \\
C & 316$\pm$22   & 1.24$\pm$0.23 &  361$\pm$15   & 1.52$\pm$0.09   \\
W & 424$\pm$57  & 2.31$\pm$0.67 & 403$\pm$128    & 1.68$\pm$0.18   \\ \hline
Mean & 352$\pm$62 & 1.79$\pm$0.54 & 387$\pm$23 & 1.75$\pm$0.27\\
\hline
\end{tabularx}
\end{minipage}
\end{table}

\begin{table}[h]
\centering
\begin{minipage}{1\columnwidth}
\caption[]{Vertical scale heights of thin disk ($\rm{h_{thin}}$) and halo ($\rm{h_{halo}}$) components of exponential fits to synchrotron emission profiles. The western (W), central (C) and eastern (E) strips have a width of 180$''$ each.}
\label{scale_heights_synchrotron}
\begin{tabularx}{1\columnwidth}{c*5{>{\centering\let\\=\tabularnewline}X}}
\hline \hline
\multirow{2}{*}{Strip}  & \multicolumn{2}{c}{5.99~GHz} & \multicolumn{2}{c}{1.57~GHz} \\
                                 &  $\rm{\bm h_{\bm t\bm h\bm i\bm n}}$ [pc] & $\rm{\bm h_{\bm h\bm a\bm l\bm o}}$ [kpc] & $\rm{\bm h_{\bm t\bm h\bm i\bm n}}$ [pc] & $\rm{\bm h_{\bm h\bm a\bm l\bm o}}$ [kpc]\\ \hline
E & 863$\pm$702   & 5.01$\pm$12.48 & 255$\pm$1097   & 1.97$\pm$0.23  \\
C & 351$\pm$14   & 1.31$\pm$0.14 &   361$\pm$12   & 1.45$\pm$0.06   \\
W & 398$\pm$116  & 1.90$\pm$0.34 &   413$\pm$291   & 1.56$\pm$0.19   \\ \hline
Mean & 375$\pm33$\tablefootmark{a} & 1.60$\pm$0.42$^{a}$ & 343$\pm$81 & 1.66$\pm$0.27\\
\hline
\end{tabularx}
\tablefoot{\tablefoottext{a}{Excluding the eastern strips at 5.99~GHz, see Sect.~\ref{section_scale_heights} for details.}}
\end{minipage}
\end{table}

\subsection{Spectral index distribution between 1.57 and 5.99~GHz}
\label{total_spix_distribution_section}

\begin{figure*}[]
\centering
\begin{subfigure}{}
\includegraphics[trim = 10mm 178mm 35mm 15mm, clip, width=1\columnwidth]{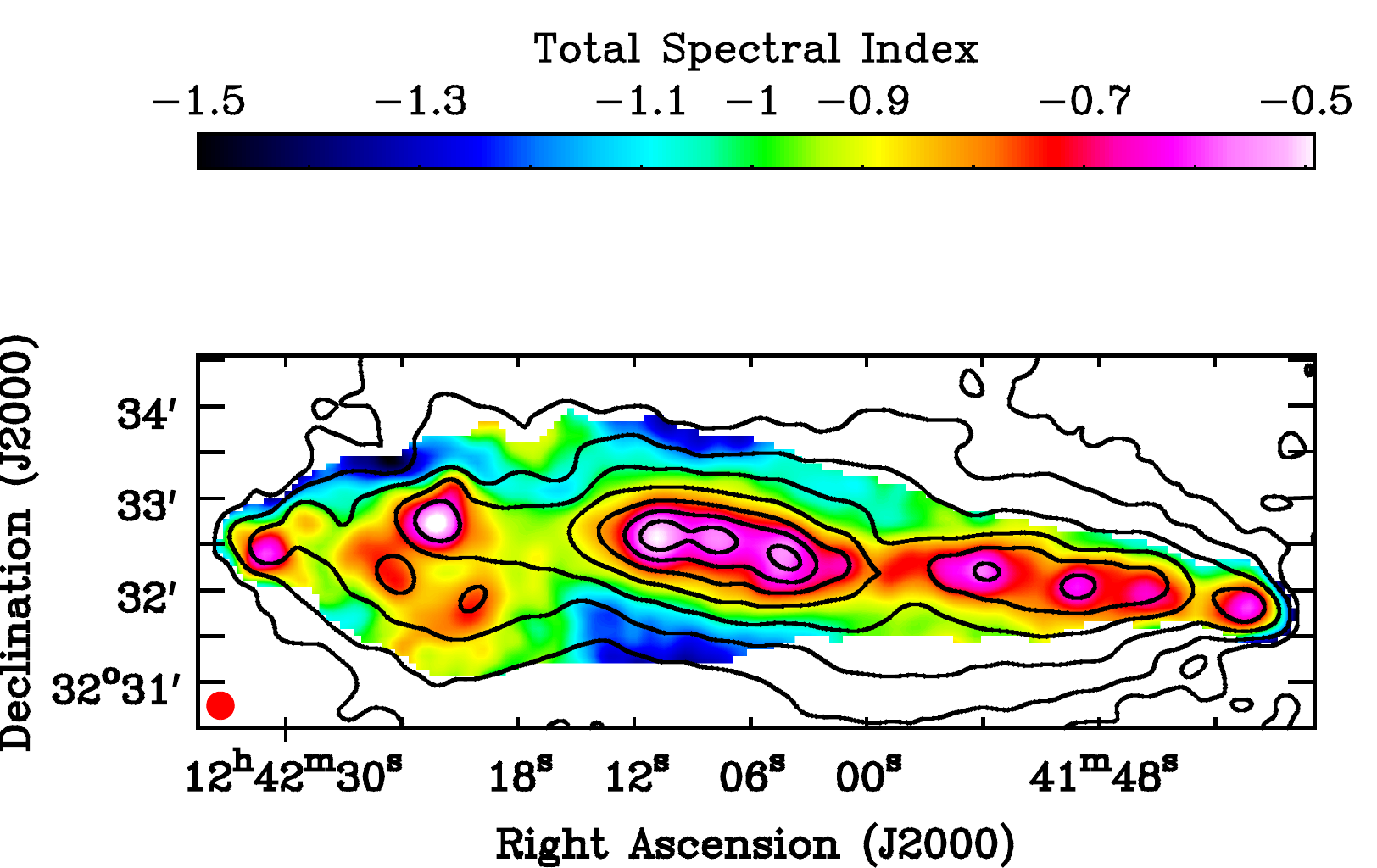}
\end{subfigure}
\begin{subfigure}{}
\includegraphics[trim = 20mm 193mm 50mm 15mm, clip, width=1.\columnwidth]{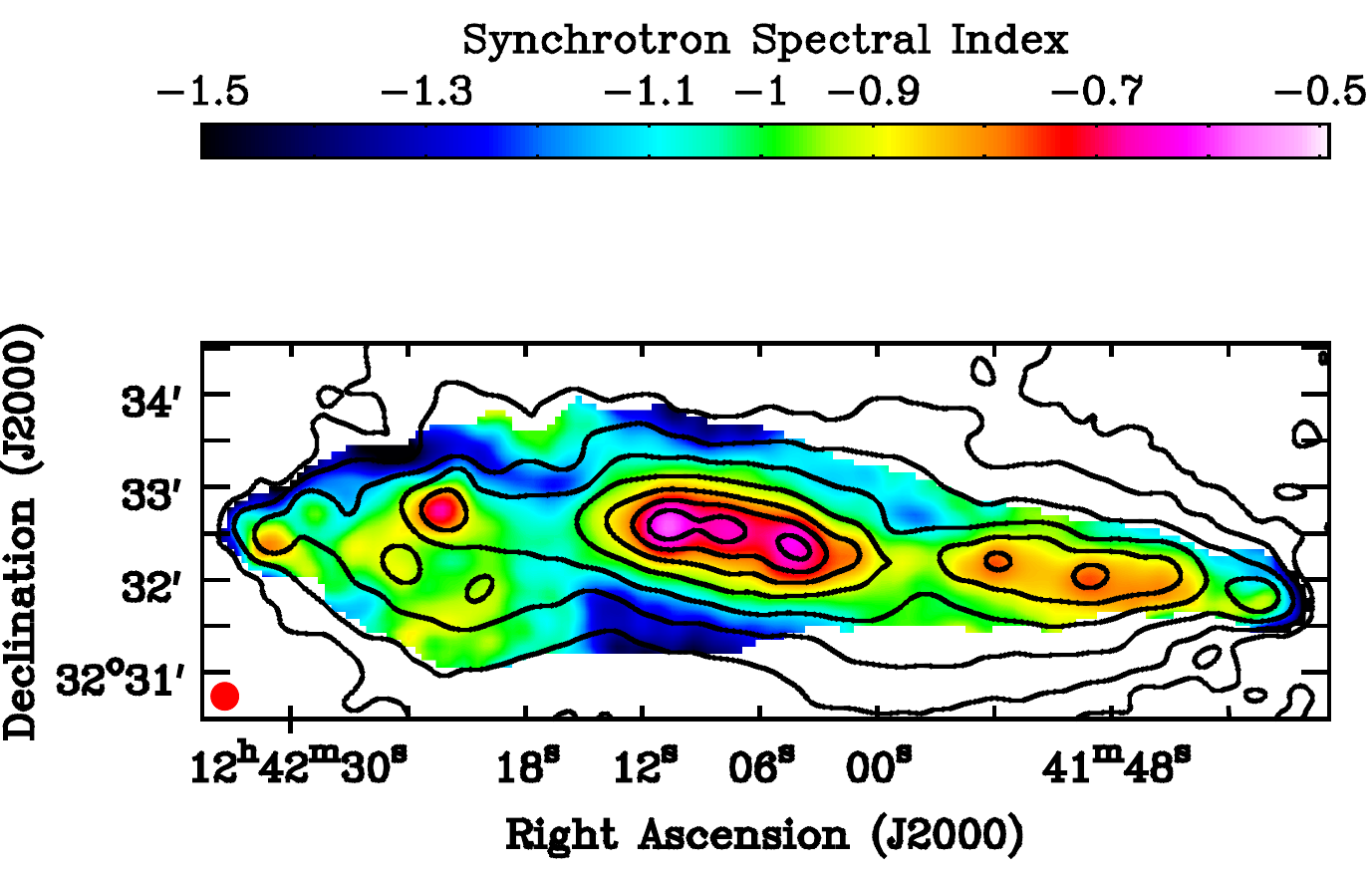}
\end{subfigure}
\caption[]{Left panel: Total spectral index distribution (in color) between 1.57 and 5.99~GHz data (VLA + Effelsberg). Right panel: Synchrotron spectral index distribution (in color) derived between 1.57 and 5.99~GHz data (VLA + Effelsberg). In both figures, values were calculated at pixels where the H$\alpha$ emission and total power are larger than 3$\sigma$. All data plotted have an angular resolution of 18$''$~FWHM. Contour levels corresponding to the 5.99~GHz total power emission (VLA + Effelsberg) are at $35~\rm{\mu Jy/beam}\times(3, 6, 12, 24, 48, 96, 192, 384)$. }
\label{spix_maps}
\end{figure*}

The distribution of the spectral indices of the total radio continuum emission ($\alpha_{\rm tot}$) computed between data at 1.57 and 5.99~GHz (VLA and Effelsberg data combined) are shown in the left panel of Figure~\ref{spix_maps}. Values were determined only above a 3$\sigma$ threshold in both maps. The largest error of the total spectral index values is 0.1 in the lowest signal-to-noise ratio regions and decreases with increasing intensity. The error in the total spectral index determination was estimated by error propagation, assuming that the dominating factor is the rms noise in the maps. Uncertainties due to flux density calibration were excluded because they affect both frequencies in the same way.

The eastern peak of the triple radio source in the central region of NGC~4631, together with the radio peak to the eastern side of the disk ($\alpha_{2000}=12^\mathrm{h}42^\mathrm{m}22^\mathrm{s}, \delta_{2000}=32^{\circ}32'50 \farcs5$), have the flattest spectrum within the galaxy with an average value of $\rm{\alpha_{tot}}=-0.53\pm0.02$. These regions are well known to coincide with H{\sc ii} complexes.

The synchrotron spectral index ($\alpha_{\rm syn}$) distribution shown in Figure~\ref{spix_maps} (right panel) was derived using the combined (VLA + Effelsberg) synchrotron emission maps at 1.57 and 5.99~GHz. Only pixels where the H$\alpha$ emission (from which the thermal emission was derived) and the total power emission are larger than 3$\sigma$, were taken into account. The largest error in the synchrotron spectral index values is 0.14 in the lowest signal-to-noise ratio regions and decreases with increasing intensity. This error includes the uncertainty related to the rms noise in the total radio maps and a 19\% (17\%) uncertainty in the 5.99 (1.57)~GHz thermal intensity estimate.

The ratio between the total and synchrotron spectral index obtained in spatially resolved regions of NGC~4631 is shown in Table~\ref{alpha_alphanth_ratio}. For convenience, we define the central region of the galaxy as the area within the fifth contour in Fig.~\ref{spix_maps} extending in right ascension between $12^\mathrm{h}42^\mathrm{m}00^\mathrm{s}$ and $12^\mathrm{h}42^\mathrm{m}18^\mathrm{s}$. As expected, the ratio $\rm{\alpha_{tot}/\alpha_{syn}}$ is below unity for all areas. The higher the thermal fraction, the steeper the synchrotron spectral index is with respect to the total spectral index. For a thermal fraction of 20\% at 5.99~GHz, we have a variation of about 12\% between the synchrotron spectral index values and the total spectral indices between 1.57 and 5.99~GHz. A similar variation of $\alpha_{\rm tot}/\alpha_{\rm syn}$ with thermal fraction is also seen in face-on galaxies \citep{Basu2012}.

\begin{table}[h!]
\centering
\begin{minipage}{1\columnwidth}
\caption[Ratio between total and synchrotron spectral index]{Ratio between total and synchrotron spectral index ($\rm{\alpha_{tot}/\alpha_{syn}}$) obtained in spatially resolved regions of NGC~4631. The thermal fractions at 5.99~GHz ($\rm{f_{th,5.99~GHz}}$) have also been included.}
\label{alpha_alphanth_ratio}
\begin{tabularx}{1\columnwidth}{c*3{>{\centering\let\\=\tabularnewline}X}}
\hline \hline
Region & $\rm{ f_{ th,5.99~GHz}}$ & $\rm{ \alpha_{ tot}/ \alpha_{ syn}}$  \\ \hline
CM~7,8,9\tablefootmark{a}                 & $55\pm10\%$  & 0.64      \\
CM~81\tablefootmark{b}    & $42\pm8\%$  & 0.72    \\
Disk\tablefootmark{c} & $24\pm5\%$ &0.85 \\  
Central region           & $20\pm4\%$ & 0.88 \\
Halo  & $17\pm4\%$ & 0.91\\
\hline
\end{tabularx}
\tablefoot{
\tablefoottext{a}{Location: $ \alpha_{2000}=12^\mathrm{h}41^\mathrm{m}40.2^\mathrm{s}, \delta_{2000}=32^{\circ}31'49''$}
\tablefoottext{b}{Location: $ \alpha_{2000}=12^\mathrm{h}42^\mathrm{m}22^\mathrm{s}, \delta_{2000}=32^{\circ}32'50 \farcs5$}
\tablefoottext{c}{Excluding the central region.}}
\end{minipage}
\end{table}

\subsection{Magnetic field strength}
\label{bfield}

The strength of the total magnetic field in NGC~4631 was estimated using the synchrotron intensity and the synchrotron spectral index distribution. Assuming equipartition between the energy densities of cosmic ray particles and magnetic field, which is valid in star-forming galaxies at scales of larger than about 1\,kpc \citep{Seta}, the magnetic field strength was derived using the revised equipartition formula presented in \cite{Beck+Krause}.

We assumed a constant ratio of proton-to-electron number densities of $\rm{K_0} \simeq 100$, and the energy index of the cosmic-ray energy spectrum is the same for protons and electrons. The pathlength $L$ was modeled as a combination of two oblate spheroidals:
$L=2R(1-(x/R)^2-(z/H)^2)^{1/2}$ (where $x$ and $z$ are the distances from the galaxy's center on the major and minor axis, respectively), one spheroidal corresponding to the disk, since most of the emission is concentrated in this region, and another spheroidal corresponding to the halo. The x-axis coincides with the major axis of the galaxy, the z-axis with the minor axis, and the y-axis represents the pathlength through NGC~4631. Since the radio emission of NGC~4631 extends to 15~kpc in radius along the major axis, we assumed the maximum pathlength through the galaxy to be twice this value. Therefore, the semi-major axes of the disk spheroid are $\rm{R = 15~kpc}$ and $\rm{H = 1.5~kpc}$, while the halo spheroid has semi-major axes of $\rm{R = 15~kpc}$ and $\rm{H = 6.3~kpc}$. In the central area of the galaxy, we estimated the largest pathlength of 30~kpc, which then decreases to about 7~kpc towards the outer edges of the disk. In the halo, the largest pathlength is $L\simeq29$~kpc close to the central region of the galaxy and then decreases to about 4~kpc towards the outer edges of the observable halo. The magnetic field strength $B$ shows only a weak dependence on pathlength, $B \varpropto L^{-1/(3 - \alpha_{\rm syn})}$ \citep{Beck+Krause}. Hence, our assumptions concerning the pathlength through NGC\,4631 will not have a large impact on the estimated total magnetic field strength.

The synchrotron spectral index distribution ($\rm{S_\nu \propto \nu^{ \alpha}}$) was derived by smoothing the synchrotron emission at 5.99 and 1.57~GHz to a resolution of 20$\farcs$5. Models of CR particle acceleration predict injection energy spectra with slopes between $-2.0$ and $-2.4$ \citep[see e.g., Fig.~3 in][]{Caprioli2011}, leading to synchrotron spectral indices between $-0.5$ and $-0.7$. Steeper spectral index values of $\rm{\alpha_{syn} < -1.1}$ probably arise due to strong energy losses (synchrotron and inverse Compton) of cosmic ray electrons (CREs), which steepens the energy spectrum of the CREs. To reduce an effect on the equipartition field strength estimate, we set $\rm{\alpha_{syn} = -1.1}$ wherever $\rm{\alpha_{syn} < -1.1}$. Furthermore, the ratio between the number densities of the relativistic protons and electrons is expected to increase in the halo. Our assumption of a constant ratio $\rm{K_0}$ leads to an underestimate of the field strengths in the halo.

\begin{figure}[h]
\centering
\includegraphics[trim = 10mm 150mm 20mm 18mm, clip, width=1\columnwidth]{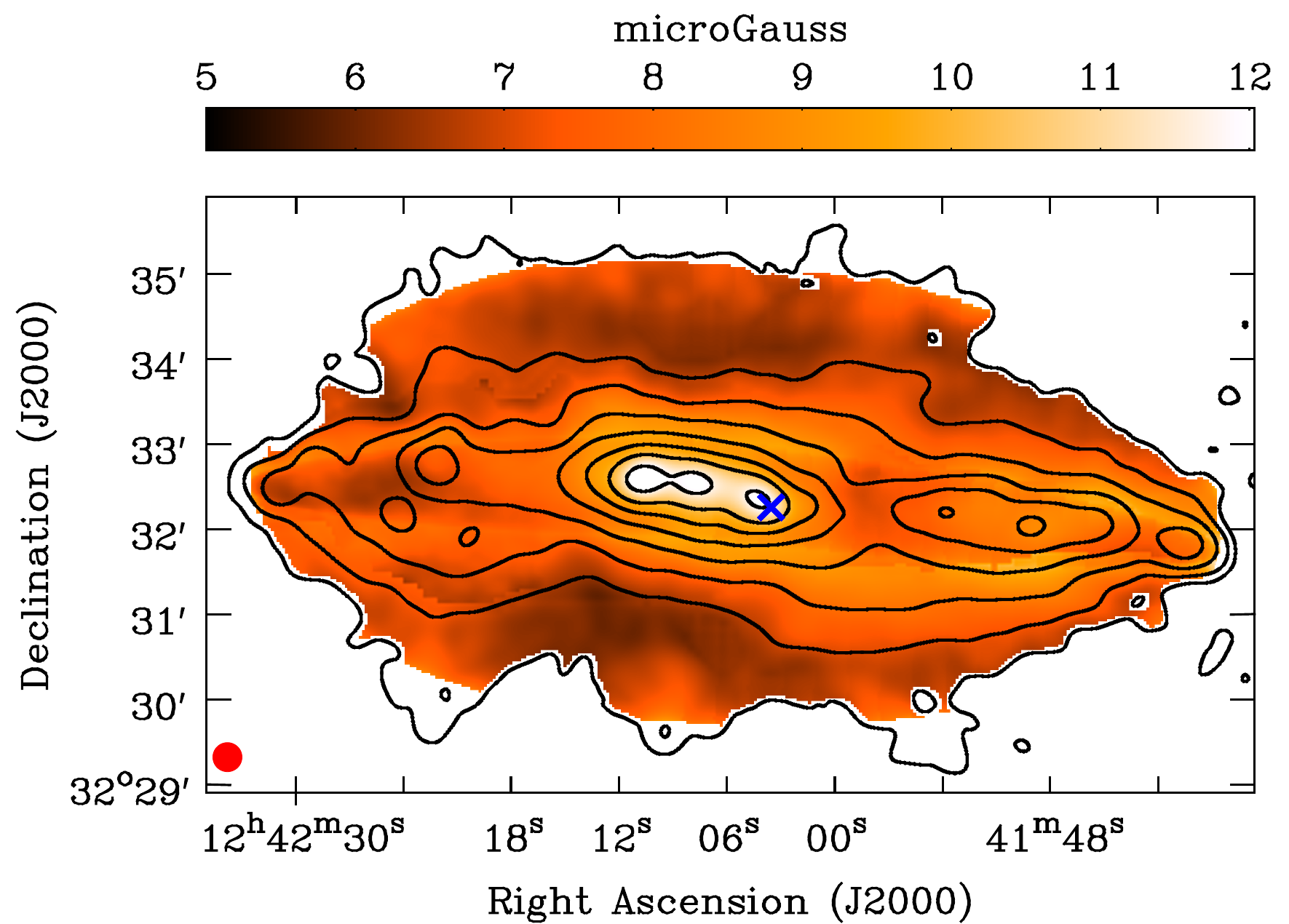}
\caption[Strength of total magnetic field]{Equipartition strength of total magnetic field in NGC~4631. The angular resolution is 20$\farcs$5~FWHM. Contours correspond to the 5.99~GHz total power emission (VLA + Effelsberg) are at $45~\rm{\mu Jy/beam}\times(3, 6, 12, 24, 48, 96, 192, 384)$. The background radio galaxy located at the central region of NGC~4631 has been marked with a blue X-symbol (see Appendix~\ref{BGRadioGalaxies}). }
\label{total_field_strength}
\end{figure}

The strength of the total magnetic field of NGC~4631 is shown in Figure~\ref{total_field_strength}. The errors are at most 50\% towards the outer parts of NGC~4631 and decrease with increasing intensity in the inner regions. The errors in the total magnetic field strengths were estimated using the Monte-Carlo method by taking into account the uncertainties of the synchrotron intensity values and the error of the synchrotron spectral indices as described in \cite{Basu2013}. The mean strength in the disk is $\rm{\langle B_{eq}\rangle} \simeq 9~\rm{\mu G}$, but in the central area of the galaxy the field strength goes up to $\rm{B_{eq}}\simeq 13~\rm{\mu G}$. In the halo, the mean field strength is lower, $\rm{\langle B_{eq}\rangle}\simeq 7~\rm{\mu G}$. 
These values are somewhat smaller than those estimated by \cite{Mora2013} with Effelsberg data at 8.35~GHz assuming a constant nonthermal spectral index. The mean magnetic field strength in the disk of NGC~4631 is comparable to that in NGC~5775 \citep{Soida2011}, but is not as strong as in NGC~253 \citep{Heesen2009_I}, which is known to have very strong star formation activity in its disk.

\section{Discussion}
\label{Discussion}

\subsection{Equipartition in the halo?}
\label{equipartition}

The magnetic field strengths shown in Fig.~\ref{total_field_strength} are computed by assuming energy density equipartition between cosmic rays and magnetic fields. The equipartition assumption was shown to hold in the local interstellar medium (ISM) of the Milky Way \citep[e.g.,][]{Boulares1990} and yields field strengths that are similar to those obtained with other methods, for instance, from the combination of radio synchrotron and $\gamma$-ray data \citep[e.g., in M~82,][]{Yoast2013}. However, equipartition does not hold on spatial scales smaller than the diffusion length of cosmic rays (see \citet{Beck2015_II} for a detailed discussion).

\citet{Heesen2018} used images of NGC~4631 at 1.37~GHz \citep{Braun2007} and at 4.86~GHz re-reducing the data by \cite{Mora2013}, to compute vertical profiles of synchrotron intensity and synchrotron spectral index, averaged over a strip of $5.8\arcmin$ width along the galaxy's major axis, and fit models of the propagation of CREs. From their model, the profile of the total magnetic field strength can be determined without the assumption of local energy equipartition between cosmic rays and magnetic fields. The profile of their best-fit model (Fig.~3d of their paper) can be described by two exponential functions with the scale heights of 0.4~kpc and 0.6~kpc in the disk above and below the plane, respectively, and of 3.2~kpc and 5.0~kpc in the northern and southern halo, respectively. To investigate whether equipartition is valid in the model of \citet{Heesen2018}, we compute the normalized synchrotron intensities, $I_{\rm syn} \propto n_{\rm CRE}\,B_\perp^{\,\,1-\alpha}\,V$ in the central region and in the halo:

\begin{equation}
\dfrac{I_{\rm syn, c}}{I_{\rm syn, h}} = \left(\dfrac{n_{\rm CRE, c}}{n_{\rm CRE, h}}\right)\, \left(\dfrac{B_{\rm c}^{\,\,1-\alpha_{\rm c}}}{B_{\rm h}^{\,\,1-\alpha_{\rm h}}}\right) \, \left(\dfrac{V_{\rm c}}{V_{\rm h}}\right),
\end{equation}
where the subscript `c' refers to values in the central region (averaged over the inner box of the central strip in Fig.~\ref{scale height}) and subscript `h' refers to the halo at 4~kpc height. $n_{\rm CRE}$ is the CRE number density. $V$ is the synchrotron emitting volume, $V = {\rm beam ~area}\times L$ where $L$ is the synchrotron pathlength. For the same beam area, $V_{\rm c}/V_{\rm h} = L_{\rm c}/L_{\rm h}$. From Sect.~\ref{bfield} we adopt $L_{\rm c} = 30$~kpc and $ L_{\rm h} = 8$~kpc.

From our data, we find $S_{\rm syn, c} \simeq 8$~mJy/beam and $S_{\rm syn, h} \simeq 0.1$~mJy/beam at 5.99~GHz and at a resolution of 18$''$. Fig.~\ref{spix_maps} (right panel) gives $\alpha_{\rm c} \simeq -0.8$ and $\alpha_{\rm h} \simeq -1.0$. Applying the equipartition assumption to the synchrotron intensities yields $B_{\rm c} \approx 13\,\mu$G and $B_{\rm h} \approx 6\,\mu$G, so that the intensity ratio of

\begin{equation}
\dfrac{8}{0.1} = \left(\dfrac{n_{\rm CRE, c}}{n_{\rm CRE, h}}\right)\, \left(\dfrac{13^{1.8}}{6^2}\right) \, \left(\dfrac{30}{8}\right),
\end{equation}
yields $n_{\rm CRE, c}/n_{\rm CRE, h} \approx 7$, that is, the CRE number density (and hence also the energy density of total CRs, assuming a constant p/e ratio $K_0$) in the central region is about $7\times$ higher than that in the halo. A similar ratio is valid for the magnetic energy densities (($13/6)^2 \approx 5$), as expected for our assumption of energy equipartition.

\citet{Heesen2018} give $S_{\rm syn, c} \simeq 10$~mJy/beam and $S_{\rm syn, h} \simeq 0.2$~mJy/beam at 4.86~GHz and at a resolution of 23$''$ (their Fig.~3a), $\alpha_{\rm c} \simeq -0.65$ and $\alpha_{\rm h} \simeq -1.2$ (their Fig.~3b), and $B_{\rm c} \simeq 13~\mu$G and $B_{\rm h} \simeq 2~\mu$G (their Fig.~3d). The intensity ratio

\begin{equation}
\dfrac{10}{0.2} = \left(\dfrac{n_{\rm CRE, c}}{n_{\rm CRE, h}}\right)\, \left(\dfrac{13^{1.65}}{2^{2.2}}\right) \, \left(\dfrac{30}{8}\right),
\end{equation}
gives $n_{\rm CRE, c}/n_{\rm CRE, h} \approx 1.0$, hence a similar number density of CREs in the center as in the halo. This seems consistent with the argument presented in \citet{Heesen2018} that ``the energy loss in the halo is therefore escape dominated, which implies that the CREs leave the halo so fast that no significant frequency dependence due to synchrotron losses can be observed.''
The ratio of magnetic energy densities between central region and halo is about $(13/2)^2 \approx 40$.

According to \citet{Heesen2018}, the cosmic ray energy density in the halo is a factor of about 40 larger than the magnetic one. If so, the cosmic rays would totally dominate the energy budget, so that the cosmic rays would escape together with the gas, and the field lines would be dragged outwards with the wind. Such a scenario seems consistent with the observed pattern of the magnetic field in the halo 
(refer to Fig. 1 in  \citealt{Mora-Partiarroyo2019_B}). However, the large absolute values of Faraday depth found in the halo (refer to Fig. 5 in  \citealt{Mora-Partiarroyo2019_B}) are in conflict with total halo fields of a few $\mu$G strength in \citet{Heesen2018}, of which only a small fraction (i.e., the line-of-sight component of the regular field) can contribute to Faraday depth (refer to Sect. 4 in \citealt{Mora-Partiarroyo2019_B}).

We believe that a drastic deviation from equipartition like in \citet{Heesen2018} is unphysical. Instead, we expect that equipartition is valid also in radio halos driven by galactic winds. Magnetic fields are required to couple the cosmic rays to the gas flow. Cosmic rays excite Alfv{\'e}n waves that in turn scatter cosmic rays and limit their propagation with respect to the gas \citep{Kulsrud1969}. On the other hand, the pressure of cosmic rays is needed to drive galactic winds \citep[e.g.,][]{Wiener2017}. An improved CRE propagation model using the new data presented in this paper will follow (Heesen et al., in prep.).

\subsection{Cosmic-ray electron energy losses and galactic winds}
\label{cosmic_ray_losses}
 
The CREs in the ISM are produced in supernova explosions and are subject to energy losses through inverse-Compton radiation (IC), synchrotron radiation, gas ionization, and bremsstrahlung. Synchrotron and IC losses steepen the spectrum towards higher frequencies, ionization losses flatten the spectrum towards lower radio frequencies, while bremsstrahlung losses do not considerably change the spectrum \citep{Pacholczyk, Longair}. We estimated the contribution of the different loss processes in NGC~4631 by determining the timescales of each of these mechanisms separately, following \cite{Heesen2009_I} for the central area of the galaxy, the disk (excluding the central region), and the halo. The values are presented in Table~\ref{lifetimes_table}. We found that synchrotron losses dominate over radiation losses by the other three processes in all three regions.

\begin{table}[h]
\centering
\begin{minipage}{1\columnwidth}
\caption[]{Cooling timescales for CREs due to synchrotron ($\tau_{\mathrm{syn}}$), inverse-Compton ($\tau_{\mathrm{IC}}$), ionization ($\tau_{\mathrm{ion}}$) and bremsstrahlung ($\tau_{\mathrm{brem}}$) losses in different regions within NGC~4631 at $\nu = 5.99~\mathrm{GHz}$. The strength of the total magnetic field perpendicular to the LOS ($\rm{B_{eq,\perp}}$), the energy ($E$) of the CREs, and the average number density of neutral gas ($\langle n\rangle$) are also included. The shortest timescales, indicating the dominant loss process, are marked in bold.}
\label{lifetimes_table}
\begin{tabularx}{1\columnwidth}{c*4{>{\centering\let\\=\tabularnewline}X}}
\hline \hline
Region & Central region & Disk\tablefootmark{a} & Halo \\ 
\hline 
$\rm{B_{eq, \perp}}$~[$\rm{\mu}$G] & 9.4 & 6.4 & 5.5 \\
$\rm{E}$~[$\rm{GeV}$] & 6.3 & 7.6 & 8.2 \\
\rm{$ \langle n \rangle$~[cm$^{-3}$]} & 1.5 & 0.8 & 0.02 \\
\rm{$\bf \tau_{\rm syn}$~[yr]} & $\bf{1.5\times 10^7}$ & $\bf{2.7\times10^7}$ & $\bf{3.3\times10^7}$ \\
\rm{$\tau_{\rm IC}$~[yr]} & 2.7$\times10^7$  & 1.4$\times10^8$    &  1.2$\times10^8$ \\
\rm{$\tau_{\rm ion}$~[yr]} & 3.6$\times10^8$ &7.9$\times10^8$ & 3.2$\times10^{10}$ \\
\rm{$\tau_{\rm brem}$~[yr]} & 2.6$\times10^7$  &4.9$\times10^7$ & 2.0$\times10^9$   \\
\hline 
\end{tabularx}
\tablefoot{\tablefoottext{a}{Excluding the central region.}}
\end{minipage}
\end{table}

\begin{figure}[h]
\centering
\includegraphics[trim = 0mm 0mm 0mm 0mm, clip,width=0.90\columnwidth]{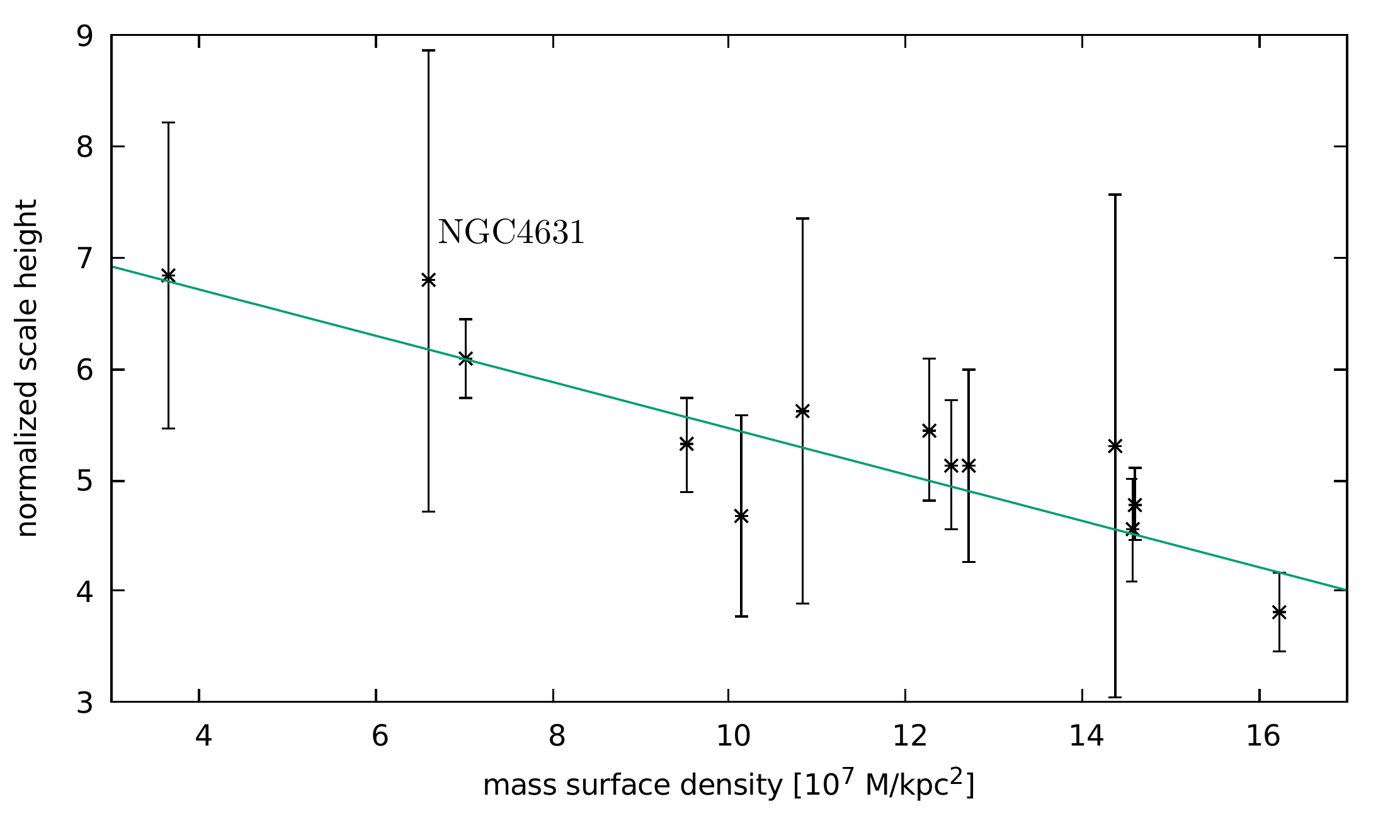}
\caption[]{Normalized scale heights at C-band vs. mass surface density of NGC~4631, together with CHANG-ES subsample of 13 galaxies from \cite{krause+2018}. The fit to the data has a slope of $-0.21 \pm 0.03$ (reduced $\chi^2 = 0.34$).}
\label{MSD}
\end{figure}

Adiabatic expansion (a galactic wind), however, may be important or even dominant in the halo, as demonstrated by \cite{krause+2018} for a sample of 13 spiral galaxies of the CHANG-ES sample. They also found that the halo scale heights at C-band and L-band depend mainly on the diameter of the radio emission along the major axis. These authors proposed to determine a normalized scale height $\tilde{h} = 100 \cdot h_{halo} / d_{\rm r}$, where $d_{\rm r}$ is the corresponding radio diameter along the major axis. We determined $\tilde{h}$ for the total radio emission of NGC~4631, resulting in $6.79\pm2.08$ at C-band and $6.35\pm1.03$ at L-band, very similar to each other. With the total mass of $5.5\times10^{10}\,$M$_{\odot}\,$ \citep{tully1988}, NGC\,4631 has a mass surface density of $6.59\times10^7\, $M$_{\odot} \, $kpc$^{-2}$. With these values, NGC~4631 fits well at both frequency bands into the linear anti-correlation of the normalized scale height with the mass surface density, as found by \cite{krause+2018}. The plot for C-band, including NGC~4631, is presented in Fig.~\ref{MSD}. The fit values for L-band are comparable.

The values for the mean scale heights in NGC~4631 are similar at C-band and L-band (see Table~\ref{scale_heights}), hence do not show any frequency dependence, as would have been expected if radiation losses (synchrotron and IC losses) dominate over propagation losses (galactic wind). We therefore, conclude that the energy loss in the halo is escape dominated, which implies that the CREs leave the halo so fast that no significant frequency dependence due to synchrotron losses can be observed. Following \cite{krause+2018}, this means that the timescale for adiabatic losses of CREs is smaller than their synchrotron lifetime within the observed distances from the galactic plane. Our estimate for the synchrotron lifetime of CREs in the halo is $t_{\rm syn} \simeq 3.3 \times 10^7\,$yr at C-band with a magnetic field strength $B = 6\, \mu$G in the outer halo. If we assume equipartition between magnetic fields and cosmic rays, the scale height of the cosmic-ray electrons $h_{\rm CRE}$ is about twice the observed synchrotron scale height $h_z \simeq 1.7$~kpc (see $ h_{halo}$ in Tables \ref{scale_heights} and \ref{scale_heights_synchrotron}). Hence, we can estimate a mean cosmic-ray bulk speed $\mathrm{v_{CRE}}$, corresponding to the velocity of a galactic wind (if we neglect cosmic ray streaming) of $\mathrm{v_{CRE}} = 2 h_z / t_{\rm esc} > 2 h_z / t_{\rm syn} \simeq 100\,$km/s. This is a lower limit because escape is probably dominated by adiabatic expansion.
 
The adiabatic loss time depends only on the reciprocal velocity gradient: $t_{\rm ad} = 3 \left(\frac{\partial {\mathrm{v}}}{\partial z}\right)^{-1}$ \citep{Heesen2009_I}. Assuming a constant increase of the CRE velocity with height, gives $\mathrm{v_{\rm wind}} = 6 h_z / t_{\rm ad} = 300\,$km/s at a height of about 3~kpc above the galactic plane. This is similar to the escape velocity from the disk, which is about 280~km/s for a rotation speed of 200~km/s \citep{Heesen2009_I}. Hence we conclude that the radio halo is escape-dominated with convective cosmic ray propagation, that is, there is a galactic wind in the halo of NGC~4631.

\section{Summary and conclusions}
\label{Conclusions}

Radio continuum observations of NGC~4631 were performed with the Karl G. Jansky Very Large Array at C-band (5.99~GHz) in the C \& D array configurations, and at L-band (1.57~GHz) in the B, C, \& D array configurations. In order to recover the large-scale emission that is missing due to the incomplete u,v sampling of the interferometer, the total intensity data were combined with single-dish Effelsberg data. Observations of NGC~4631 with the Effelsberg 100-m telescope were performed at 1.42 and 4.85~GHz. We separated the thermal and synchrotron components of the total radio emission by estimating the thermal contribution through the extinction-corrected H$\alpha$ emission. The H$\alpha$ radiation was corrected for extinction using a linear combination of the observed H$\alpha$ and 24~$\mu$m data, using different calibration factors for the inner and outer emission. Our conclusions are as follow:

\begin{itemize}
\item The integrated thermal flux-density of NGC~4631 is 61$\pm$13~mJy at 5.99~GHz and 70$\pm$13~mJy at 1.57~GHz. This corresponds to a global thermal fraction at 5.99 (1.57)~GHz of 14$\pm$3\% (5.4$\pm$1.1\%).

\item The power-law spectral index of the integrated synchrotron emission is $-0.84 \pm 0.02$, without signatures of a spectral break or a cutoff.

\item The mean radio scale height of the halo at 5.99 (1.57)~GHz is $1.8\pm0.5$~kpc ($1.8\pm0.3$~kpc). The scale height of the disk observed at 5.99 (1.57)~GHz is $350\pm60$~pc ($390\pm20$~pc). These values do not deviate strongly from the values for synchrotron emission. The normalized scale height for the halo in NGC~4631 is $6.8\pm2.1$ ($6.4\pm1.0$) at 5.99 (1.57)~GHz, and perfectly fits into the linear anti-correlation of the normalized scale height with the mass surface density, as found by \cite{krause+2018}.

\item The halo scale heights do not show a frequency dependence as expected for a synchrotron loss-dominated halo. We estimated that the radio halo is escape-dominated with convective cosmic ray propagation, and conclude that there is a galactic wind in the halo of NGC~4631.

\item The total equipartition magnetic field strength of NGC~4631 using the synchrotron spectral index map is estimated to be $\rm{\langle B_{eq}\rangle} \simeq 9~\rm{\mu G}$ in the disk and $\rm{\langle B_{eq}\rangle}~\simeq 7~\rm{\mu G}$ in the halo, which is somewhat smaller than values estimated previously assuming a constant spectral index.

\item We analyzed the background radio galaxy to the southwest of NGC~4631. Its spectral index distribution derived with 1.57 and 5.99~GHz data is flat in the outskirts of the lobes and steepens towards the core. This is typical of FR~II radio galaxies and is consistent with the backflow model. The background radio galaxy has an average nonthermal degree of polarization of 7\% at 1.57~GHz and its Faraday depth is rather uniform over the two lobes with an average value of $-21 \pm 13~\rm{rad / m^{2}}$. Its intrinsic magnetic field is orientated parallel to the radio galaxy\textquoteright s main axis.
\end{itemize}

\begin{acknowledgements}
We thank Fatemeh Tabatabaei for useful recommendations concerning the thermal and nonthermal separation. This work is partly based on observations made with the Spitzer Space Telescope, which is operated by the Jet Propulsion Laboratory, California Institute of Technology under a contract with NASA. AB acknowledges financial support by the German Federal Ministry of Education and Research (BMBF) under grant 05A17PB1 (Verbundprojekt D-MeerKAT). RAMW acknowledges support from NSF Grant AST-1615594.

\end{acknowledgements}

\bibliographystyle{aa}
\footnotesize
\bibliography{References_NGC4631_AA_2017} 

\begin{appendix}

\section{Background radio galaxies}
\label{BGRadioGalaxies}

In our data we identified two radio galaxies in the vicinity of NGC~4631. One of them is located close to the center of the galaxy, at coordinates $\alpha_{2000}= 12^\mathrm{h}42^\mathrm{m}3.5^\mathrm{s}, \delta_{2000}=32^{\circ}32'16''$ (Figure~\ref{total_field_strength}). Its inner components are not resolved in our data. We estimated the spectral index of the eastern lobe to be $-0.67\pm0.01$, while the western lobe has a slightly flatter spectral index of $-0.65\pm0.01$. \cite{Golla1999} observed the central region of NGC~4631 with the VLA in the A- \& B-configuration achieving a resolution of 0$\farcs$35. At this resolution the object is resolved into at least 3 components, which are well aligned along a single axis.

\begin{figure}[h]{}
        \includegraphics[trim = 12mm 145mm 40mm 18mm, clip,width=1\columnwidth]{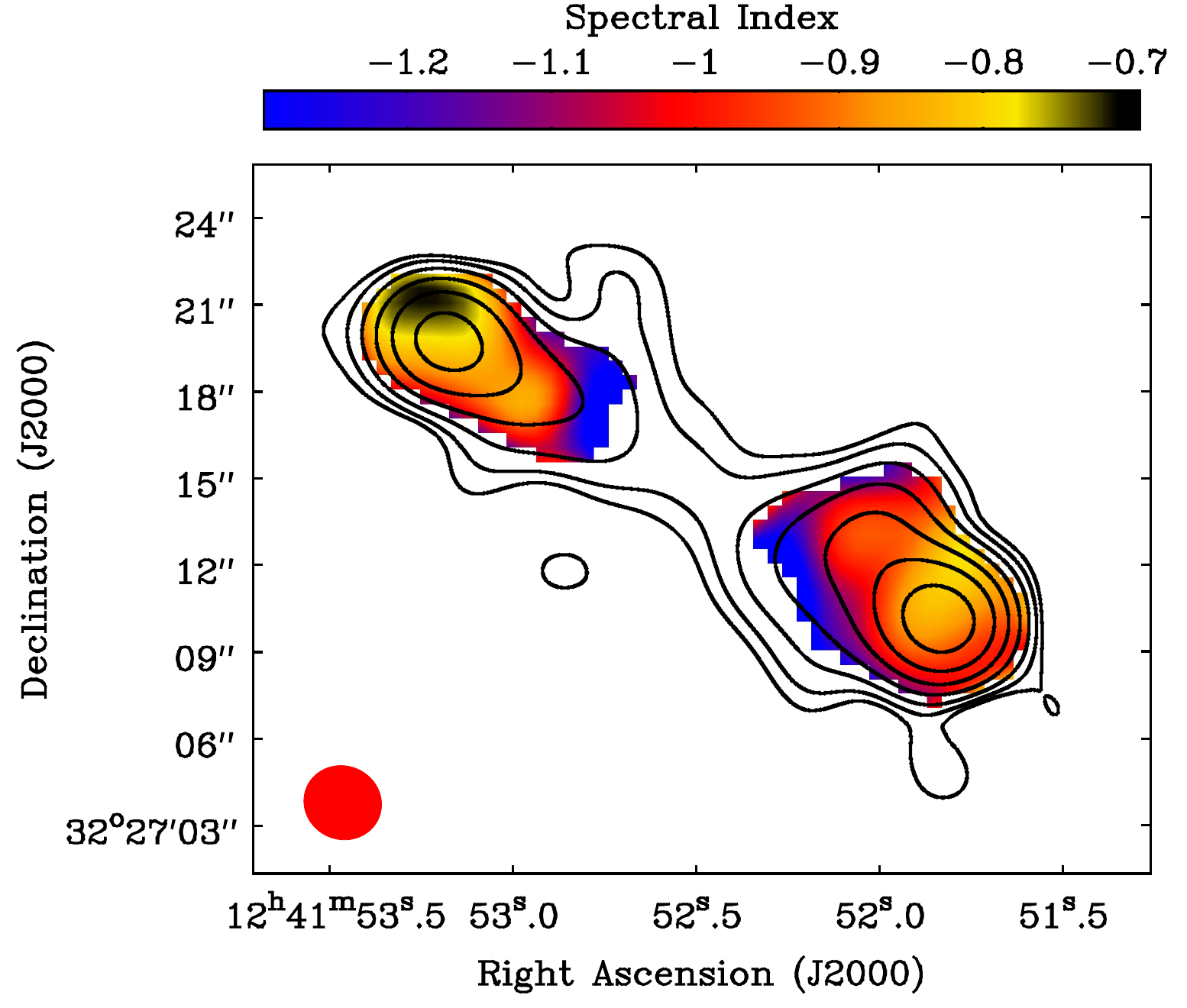}
        \caption[]{Background radio galaxy to southwest of NGC~4631. The colorscale represents the total spectral index distribution between C-band (5-6~GHz;  VLA C-configuration) and L-band (VLA B-configuration) data. The angular resolution is $2\farcs7 \times 2\farcs5$~FWHM. Contour levels corresponding to the total L-band emission (VLA B-configuration) are at $34~\rm{\mu Jy/beam}\times(3, 6, 12, 24, 48, 96, 192, 384)$.}
\label{Southern_radio_galaxy_spix}
\end{figure}

The other background radio galaxy is the bright source to the southwest of NGC~4631 at $\alpha_{2000}= 12^\mathrm{h}41^\mathrm{m}52.3^\mathrm{s}, \delta_{2000}=32^{\circ}27'14''$ (see Figure~\ref{total_radio_over_HI}). In order to fully observe the galaxy within the C-band primary beam, the data taken in the C array configuration was imaged with only half of the bandwidth (5--6~GHz), thus resulting in a larger primary beam. We present the first high-resolution analysis of this source. Its spectral index distribution determined above the 5$\sigma$ threshold is shown in Figure~\ref{Southern_radio_galaxy_spix}.

The two radio lobes can be clearly seen in Figure \ref{Southern_radio_galaxy_spix}, however, we could not detect a compact core in our radio data. The spectrum in the northern lobe steepens from -0.71$\pm$0.01 in the outer parts to -1.35$\pm$0.18 towards the nucleus of the galaxy. The spectrum in the southern lobe also steepens towards the core from -0.78$\pm$0.04 in the extremities to -1.45$\pm$0.15 closer to the center. This spectral index distribution is typical of FR~II radio galaxies \citep{Fanaroff1974} and is consistent with the backflow model of synchrotron-emitting relativistic electron plasma \citep{Leahy1984, Leahy1989}. According to this model, electrons are reaccelerated in the hotspots that give rise to a flat spectral index. The material then flows back towards the nucleus due to the pressure gradient created by the AGN jet, so that the emission closer to the core is composed of an older population of CREs which make the radio spectrum steeper. This process has also been observed in other radio galaxies, for example, \cite{Lara2000, Tamhane2015}. The flattest feature in the spectral index distribution of the northern lobe suggests the existence of a faint hotspot at the end of the lobe, while a similar feature is not visible in the southern lobe.

At 20\farcs5 resolution, this background radio galaxy appears as a point-like source with an average nonthermal degree of polarization of $~7\%$ at L-band. The Faraday depth (from L-band data) is rather uniform over the two lobes, with an average value of $-21\pm13~\rm{rad / m^{2}}$ (for details concerning the polarization data refer to \citealt{Mora-Partiarroyo2019_B}).  
The observed Faraday depth may originate in the Galactic foreground or in the faint halo of NGC~4631.

A plausible candidate for the core of this radio galaxy is CXO~J124152.3+322714, an X-ray source at position $\alpha_{2000}= 12^\mathrm{h}41^\mathrm{m}52.346^\mathrm{s}, \delta_{2000}=32^{\circ}27'14.33''$ \citep{Mineo2012}. The radio galaxy is detected at 1.4~GHz in FIRST \citep{White1997} and it has also been detected in the VLA Sky
Survey (VLASS)\footnote{https://science.nrao.edu/science/surveys/vlass} in the 2-4~GHz band at a 2\farcs5 resolution. Its redshift is unknown.

\end{appendix}

\end{document}